\documentclass[pra,twocolumn,showpacs,superscriptaddress,floatfix, nofootinbib]{revtex4-1}
\usepackage[caption=false]{subfig}
\usepackage{amsmath,graphicx}
\usepackage{epsfig}
\usepackage{amsmath,amssymb,mathrsfs,esint}
\usepackage{gensymb}
\usepackage[bookmarks=true,
   colorlinks=true,
   linkcolor=blue,
   urlcolor=blue,
   citecolor=blue,
   bookmarks=true,
   hyperindex=true
]{hyperref}
\usepackage{natbib}
\usepackage{relsize}
\usepackage{float}
\usepackage{dsfont}

\usepackage{bm}

\usepackage{soul}

\newcommand{\be}{\begin{equation}}
\newcommand{\ee}{\end{equation}}

\newcommand{\beq}{\begin{eqnarray}}
\newcommand{\eeq}{\end{eqnarray}}

\def\H1{\widehat{H}_1}

\newcommand{\ket}[1]{\left| #1 \right>}

\renewcommand{\vec}[1]{ \left|\left. #1 \right\rangle\! \right\rangle }

\renewcommand{\Re}[1]{ \text{Re} \left\{ #1\right\} }
\renewcommand{\Im}[1]{ \text{Im} \left\{ #1\right\} }

\begin{document}

\setstcolor{red}

\title{Lie-algebraic approach to one-dimensional translationally-invariant free-fermionic dissipative systems}

\author{L.R. Bakker}
\affiliation{Institute for Theoretical Physics Amsterdam, Universiteit van Amsterdam, Amsterdam 1098 XH, The Netherlands}
\affiliation{Russian Quantum Center, Skolkovo, Moscow 143025, Russia}

\author{V.I. Yashin}
\affiliation{Russian Quantum Center, Skolkovo, Moscow 143025, Russia}
\affiliation{Moscow Institute of Physics and Technology, Dolgoprudny, Moscow Region 141700, Russia}

\author{D.V. Kurlov}
\affiliation{Russian Quantum Center, Skolkovo, Moscow 143025, Russia}

\author{A.K. Fedorov}
\affiliation{Russian Quantum Center, Skolkovo, Moscow 143025, Russia}
\affiliation{Moscow Institute of Physics and Technology, Dolgoprudny, Moscow Region 141700, Russia}

\author{V. Gritsev}
\affiliation{Institute for Theoretical Physics Amsterdam, Universiteit van Amsterdam, Amsterdam 1098 XH, The Netherlands}
\affiliation{Russian Quantum Center, Skolkovo, Moscow 143025, Russia}

\begin{abstract}
We study dissipative translationally-invariant free fermionic theories with quadratic Liouvillians. 
Using a Lie-algebraic approach we solve the Lindblad equation and find the density matrix at all times for arbitrary time dependence of the Liouvillian. 
We then investigate the Liouvillian spectral properties and derive a generic criterion for the closure of the dissipative gap, which is believed to be linked with non-equilibrium dissipative phase transitions. 
We illustrate our findings with a few exotic examples. 
Particularly, we show the presence of gapless modes with a linear spectrum for fermions with long-range hopping, which might be related to non-unitary conformal field theories.
The predicted effects can be probed in experiments with ultracold atomic and quantum-optical systems using currently available experimental facilities. 
\end{abstract}

\maketitle

\section{Introduction}

Recent progress in the implementation of controllable quantum systems has opened novel opportunities for studying complex many-body dynamics~\cite{Lukin2017,Monroe2017,Martinis2018,Blatt2018,Trotzky2012,Mazurenko2017,Lukin2019}.
The interplay between driving, dissipation, and many-body effects drastically changes the nature of many-body regimes in these systems~\cite{Heyl2018}. 
This opens fascinating prospects for providing insights about the properties of quantum matter and generating exotic quantum phases~\cite{KrausZoller2008,Cirac2009,Diehl2010}.
Moreover, driven-dissipative many-body systems are ideal platforms for studying non-equilibrium phase transitions that are far less understood as compared to their equilibrium counterparts~\cite{Heyl2018,Drummond1980,KrausZoller2008,Cirac2009,Prosen2008,Diehl2008,Diehl2010,DallaTorre2010}. 
At the same time, the implementation of controllable dynamics of quantum many-body systems is key for the realization of quantum computing algorithms~\cite{Cirac2009}.
Clearly any quantum computational protocol requires external driving and quantum computing devices are prone to dissipation (decoherence). 
Consequently, understanding the properties of driven-dissipative models plays a crucial role in exploring the potential of noisy intermediate-scale quantum (NISQ) devices.  
This puts the research on driven-dissipative quantum dynamics on the forefront. 

However, such a rich variety of appearing physical phenomena in driven-dissipative systems requires a proper description, which remains a challenge.
Under the assumption of Markovianity, the time evolution of dissipative quantum systems is described by the Lindblad equation~\cite{Lindblad1976, Gorini1976} of the following form:
\be\label{Lindblad_eq}
	\frac{d\rho}{dt} = - i[{\cal H},\rho] + {\cal D}\left[ \rho \right] \equiv {\cal L}\rho,
\ee
where $\rho$ is the density matrix, ${\cal L}$ is the Liouvillian superoperator, ${\cal H}$ is the Hamiltonian responsible for the unitary time evolution, and ${\cal D}$ is the dissipator governing the nonunitary evolution.
Eq.~(\ref{Lindblad_eq}) has a formal solution in terms of the time-ordered exponential:
\be \label{rho_formal}
	\rho(t) = {\cal T} \exp\left\{ \int_0^t \,d\tau {\cal L}(\tau) \right\} \rho(0).
\ee
In many cases of physical interest the Liouvillian has an explicit time dependence. For instance, when some parameters are modulated. 
The formal solution in Eq.~(\ref{rho_formal}) is then very hard to deal with.

A remarkable simplification to Eq.~(\ref{rho_formal}) arises when different terms in the Liouvillian can be identified with different (possibly non-commuting) elements of a certain Lie algebra acting in the space of density matrices (superoperator Lie algebra)~\cite{Ringel2012}. 
A solution of the Lindblad equation for the density matrix can then be converted into a product of {\it ordinary} exponentials for an arbitrary Liouvillian time-dependence using the machinery of the theory of Lie groups \cite{Wei1963, Wei1964, Charzyski2013, Ringel_2013}. 
Nevertheless, except for a few examples, this approach has not been extensively used even for quite simple models \cite{Galitski_2011, Bola_os_2015, Markovich_2017, De_Nicola_2019, Scopa_2018, Scopa_2019, De_Nicola_2020, Vernier2020}. 
Thus, the problem of the extension of this approach to relevant quantum models is of significant importance.
We note that theoretical investigation on driven-dissipative many-body quantum dynamics has been supported by a number of experimental proposals.
Examples include cold atom setups~\cite{Diehl2008,Cirac2009,Zoller2011,Baranov2012,Zoller2010,Zoller2012} and quantum optical systems~\cite{Keeling2012} (e.g., arrays of identical nonlinear cavities coupled via photon tunneling).
Recent findings related to these proposals include the analysis of topological effects, which are induced/influenced by dissipation~\cite{Zoller2010,Zoller2012} (in particular, in the absence of the unitary dynamics~\cite{Zoller2012}). 

In this work, we study one-dimensional (1D) dissipative translationally-invariant free fermionic theories with quadratic Liouvillians.
This is one of the simplest yet experimentally relevant quantum system that admits the Lie-algebraic treatment. 
Although this system has been studied previously~(see, e.g., \cite{Heyl2018} and references therein), some of the important aspects still require more detailed investigation.
In particular, we are focused on the investigation of the Liouvillian spectral properties and derive a general criterion for the closure of the dissipative gap, which is believed to be linked with nonequilibrium dissipative phase transitions. 
This simple model allows us to gain insight into a variety of interesting regimes.
We provide few examples of such regimes: (i) the presence of gapless modes with a linear spectrum for fermions with a long-range hopping, which might be related to non-unitary conformal field theories;
(ii) non-monotonic roton-like spectrum closure, which is promising for the realization of quantum computing algorithms in the presence of noise. 

Our work is organized as follows.
In Sec.~\ref{S:dynamical_symmetry}, we discuss the aforementioned Lie-algebraic approach in more detail.
We apply this algebraic approach to a generic one-dimensional translationally-invariant quadratic fermionic Liouvillian and find the density matrix at all times for an arbitrary time dependence of the Liouvillian.
In Sec.~\ref{S:Liouvillian}, we consider the Liouvillian spectral properties and derive a generic criterion for the closure of the dissipative gap.
We specifically analyze a few exotic regimes of the spectrum closure.
We conclude in Sec.~\ref{S:Conclusion}.

\section{Driven-dissipative fermions}\label{S:dynamical_symmetry}

\subsection{Lie-algebraic approach: general case} \label{S:alg_general}

Let us consider a Liouvillian that can be expressed as a linear combination of the generators~$\mathfrak{g}_j$ of some Lie algebra ${\cal A}$:
\be \label{lin_comb_gen}
	{\cal L}(t) = \sum_j \lambda_j(t) \mathfrak{g}_j.
\ee
The corresponding Lie group $G$ is then nothing else but the dynamical semi-group that governs the time evolution. 
Thus, the time-ordered exponential in Eq.~(\ref{rho_formal}), being an element in the dynamical semi-group, can be written as a product of elements in $G$. 
In other words, the density matrix at all times can be written as
\be \label{rho_anzats}
	\rho(t) = \prod_j e^{a_j(t) \mathfrak{g}_j} \rho(0),
\ee
where $a_j(t)$ are yet unknown time-dependent $c$-numbers. The latter can be found by the following algorithm. One starts by simply using Eq.~(\ref{rho_anzats}) as an Ansatz for the solution and plugging it into Eq.~(\ref{Lindblad_eq}). Carrying out the time derivative in Eq.~(\ref{rho_anzats}) and repeatedly using the identity
\be \label{Ad_action}
	e^x y = \left(e^{\text{ad}_x} y \right) e^x,
\ee
where $x,y \in {\cal A}$ and $\text{ad}_x \; \cdot \equiv [x, \cdot \,]$, one brings $\dot\rho$ to the form of Eq.~(\ref{Lindblad_eq}), i.e., 
$\dot \rho(t) = \tilde{\cal L} \rho(t)$, where $\rho(t)$ is given by Eq.~(\ref{rho_anzats}) and $\tilde{\cal L}$ is again some linear combination of the generators ${\mathfrak g}_i$, with the coefficients now depending on $a_j$ and $\dot a_j$. 
Then, requiring that $\tilde{\cal L}$ coincides with the Liouvillian in Eq.~(\ref{lin_comb_gen}), one obtains a system of coupled ordinary (usually nonlinear) differential equations for the functions $a_j$(t). 
For ${\mathfrak g}_j$ in a closed Lie algebra, this procedure is guaranteed to work. 
In cases where $\mathfrak{g}_j$ generate a more complicated algebra, e.g., polynomial, this is not necessarily true.  
Note that the ordering of the various exponentials in Eq.~(\ref{rho_anzats}) can be arbitrary. 
However, the resulting differential equations depend on the ordering. 
In some cases it is possible to find an ordering that guarantees the simplest functional form for the system of differential equations, e.g., as in Ref.~\cite{Charzyski2013}.

The crucial advantage of the outlined algebraic approach is that it can universally deal with any initial condition~$\rho(0)$ and any time dependence of the coefficients in the Liouvillian. 
Below we are going to apply this approach to a translationally-invariant free fermionic model.

\subsection{Lie-algebraic treatment for dissipative fermions}\label{S:Dissipative}

We consider the most generic one-dimensional translationally-invariant free fermionic Hamiltonian:
\be \label{H_real_space}
	{\cal H} = -\sum_{j} \sum_{n\geq1} \left( t_n c_j^{\dag} c_{j+n} + \gamma_n c_{j}c_{j+n} + \text{H.c.} \right) - \mu \sum_{j} c_j^{\dag}c_j,
\ee
where $c_j$ and $c_j^{\dag}$ are fermionic annihilation and creation operators, $\mu$ is the chemical potential, and the complex parameters $t_n$ and $\gamma_n$ are the hopping and $p$-wave pairing amplitudes, correspondingly. 
One can see that in the case of the nearest neighbour hopping and pairing, Hamiltonian~\eqref{H_real_space} reduces to the well-known Kitaev model~\cite{Kitaev2001}. 
Under the periodic boundary conditions, the Hamiltonian (\ref{H_real_space}) in momentum space reads as
\be \label{H_k}
	{\cal H} = -\sum_{k\in \text{BZ}} \left( \xi_k c_k^{\dag} c_k + i \Delta_k^* c_{-k}^{\dag} c_{k}^{\dag} - i \Delta_k c_{k} c_{-k}  \right),
\ee
where the summation is over momenta inside the Brillouin zone, $-\pi \leq k \leq \pi$, and we denoted
\be \label{xi_Delta_k}
\begin{aligned}
	\xi_k &= \mu + 2\sum_{n\geq1}|t_n| \cos \left( k n + \arg t_n \right),\\
	\Delta_k &= \sum_{n\geq1}\gamma_n \sin kn.
\end{aligned}
\ee
Note that $\xi_k \in {\mathbb R}$, but $\xi_{-k} \neq \xi_k$, unless all $t_n$ are real. Also, one always has $\Delta_{-k} = - \Delta_{k}$ and in general $\Delta_k \in {\mathbb C}$.  However, it can always be factorized as
\be \label{factorizable_Delta_k}
	\Delta_k = \sin k\, | \tilde \Delta_k| e^{i \arg \tilde \Delta_k},
\ee
where $\tilde \Delta_{-k} = \tilde \Delta_k \in{\mathbb C}$, and the phase of $\tilde \Delta_k$ can be removed from the Hamiltonian by a gauge transformation
\be \label{gauge_transformation}
	c_k = \tilde c_k\, e^{-i \arg \tilde \Delta_k /2} .
\ee
However, the phase will reappear in the coefficients of the Liouvillian.

The dissipation is described by the dissipator
\be \label{dissipator_general_form}
	{\cal D}\left[ \rho \right] = \sum_{j} \left( L_j \rho L_j^{\dag} - \frac{1}{2} \left\{ L_j^{\dag} L_j, \rho \right\} \right),
\ee
where $\left\{ \cdot, \cdot \right\} $ is the anticommutator and $L_j$ is the jump operator whose explicit form depends on the specific type of the dissipation process.

Let us take the jump operator in the most general form that is compatible with translational invariance:
\be \label{Jump_op_real_space}
	L_j = \sum_{n} \left( u_{j-n} c_n + v_{j-n} c_n^{\dag} \right),
\ee
where $u_x$ and $v_x$ are arbitrary functions. The jump operator then has the following Fourier components:
\be \label{Jump_op_k_space}
	L_k = u_k c_k + v_k c^{\dag}_{-k},
\ee
where $u_k$ and $v_k$ are the Fourier components of $u_x$ and $v_x$, correspondingly.

We then write the dissipator~(\ref{dissipator_general_form}) in momentum space as ${\cal D}[\rho] = \sum_k {\cal D}_k $, where

\be \label{D_k}
\begin{aligned}
	{\cal D}_k&[\rho] = |u_k|^2\, c_k\, \rho\, c_k^{\dag} + |v_k|^2\, c^{\dag}_{-k}\,\rho\, c_{-k} \\
		&+u_k v_k^* c_k \rho c_{-k} + u_k^* v_k c^{\dag}_{-k} \rho c^{\dag}_k - \frac{1}{2} \left\{ L^{\dag}_k L_k, \rho \right\},
\end{aligned}
\ee
with
\be \label{LdagL_k}
\begin{aligned}
	L^{\dag}_k L_k &= |u_k|^2 c^{\dag}_k c_k + |v_k|^2 \left( \mathds{1} - c^{\dag}_{-k} c_{-k} \right)\\
	&+u_k v_k^* c_{-k} c_{k} + u_k^* v_k c^{\dag}_k c^{\dag}_{-k}.
\end{aligned}
\ee

Thus, in momentum space the Liouvillian reads as
\be \label{Liouvillian_sum_k}
	{\cal L} = \frac{1}{2} \sum_{ k \in \text{BZ}} {\cal L}_k,
\ee
where the component  ${\cal L}_k$ is given by
\begin{widetext}
\be \label{Liouvillian_gens}
\begin{aligned}
	{\cal L}_k =& \;  i \xi_k \left( X_1 - X_2 \right) + i \xi_{-k} \left(X_3 - X_4 \right)
	+ 2 \Delta_k^* \left(  X_5 - X_6 \right) - 2 \Delta_k \left( X_7 - X_8 \right) \\
	&-\frac{1}{2}\Bigl[ \mathfrak{a}_{k} \bigl( X_1 + X_2 \bigr)
	+ \mathfrak{a}_{-k}  \bigl( X_3 + X_4 \bigr)
	+\mathfrak{b}_{k}  \left( X_5 + X_6 \right)
	+ \mathfrak{b}_{k}^* \left( X_7 + X_8 \right)
	+ \Lambda_k \mathds{1} \Bigr] \\
	&+\mathfrak{c}_{k} X_9 + \mathfrak{c}_{-k} X_{10}
	+ \mathfrak{d}_{k} X_{11} + \mathfrak{d}_{k}^* X_{12}
	+ \mathfrak{d}_{-k} X_{13} + \mathfrak{d}_{-k}^* X_{14}
	+\mathfrak{e}_{k} X_{15} + \mathfrak{e}_{-k} X_{16}  ,\\
\end{aligned}
\ee
\end{widetext}
and for brevity we denoted
\be \label{aux_functions_abcde}
\begin{aligned}
	&\mathfrak{a}_{k} =   |u_{k}|^2 - |v_{-k}|^2, \quad
	\mathfrak{b}_{k} =  u^*_{k}v_{k} - u^*_{-k}v_{-k}, \\
	&\mathfrak{c}_{k}  = |u_{k}|^2, \quad
	\mathfrak{d}_{k} = u_k v_k^*,\quad
	\mathfrak{e}_{k} = |v_k|^2.
\end{aligned}
\ee
In Eq.~(\ref{Liouvillian_gens}) we also introduced the function
\be \label{Lambda_k}
	\Lambda_k =  |u_{k}|^2 + |v_k|^2 +  |u_{-k}|^2 + |v_{-k}|^2
\ee
and a set of superoperators $X_j$, which act on the density matrix $\rho$ in the following way:
\be \label{superops_def}
\begin{aligned}
&X_{1,3} \, \rho=  \left( n_{\pm k} -\frac{1}{2} \right) \rho,   &&
X_{2,4} \, \rho= \rho \left(n_{\pm k} -\frac{1}{2} \right),   &&\\
&X_5 \, \rho= c_k^{\dagger} \, c_{-k}^{\dagger} \, \rho,   &&X_6 \, \rho= \rho \,  c_k^{\dagger } \, c_{-k}^{\dagger},\\
&X_7 \, \rho= c_{-k}\, c_{k}\,\rho,   &&X_8 \, \rho= \rho\, c_{-k}\, c_{k},\\
&X_9 \, \rho= c_k\, \rho\, c_k^{\dagger},   &&X_{10} \, \rho= c_{-k}\,\rho\, c_{-k}^{\dagger},\\
&X_{11} \, \rho= c_k\,\rho \,c_{-k},   &&X_{12} \, \rho= c_{-k}^{\dagger}\,\rho\, c_k^{\dagger},\\
&X_{13} \, \rho= c_{-k}\,\rho\, c_k,   &&X_{14} \, \rho= c_k^{\dagger}\,\rho\, c_{-k}^{\dagger},\\
&X_{15} \, \rho= c_{-k}^{\dagger}\,\rho\, c_{-k},   &&X_{16} \, \rho= c_k^{\dagger}\,\rho\, c_k,\\
\end{aligned}
\ee
where $n_q = c_q^{\dag}c_q$.

Let us note at this point that the gauge transformation~(\ref{gauge_transformation}) leads to
\be
	u_k \to u_k e^{-i \arg \tilde \Delta_k /2} , \qquad v_k \to v_k e^{i \arg \tilde \Delta_k /2}.
\ee
Therefore, by removing the phase of the pairing amplitude $\Delta_k$ from the Hamiltonian~(\ref{H_k}), we reintroduce it in the dissipator part of the Liouvillian via the functions $\mathfrak{b}_{k}$ and $\mathfrak{d}_{k}$ from Eq.~(\ref{aux_functions_abcde}).

Defining the commutator for two superoperators $X_i$ and~$X_j$ as
\be
	[X_i, X_j]\,\rho = X_i (X_j\rho) - X_j (X_i \rho),\label{eq:superopcomm}
\ee
one can show that the superoperators in Eq.~(\ref{superops_def}) form a closed semi-simple Lie algebra isomorphic to $\mathfrak{u}(1)\otimes \mathfrak{sl}(4,{\mathbb C})$. 
Their commutation relations are given in Appendix~\ref{A:algebra}. Since Liouvillian (\ref{Liouvillian_gens}) is linear in the generators (\ref{superops_def}), we can use the algebraic approach discussed in subsection \ref{S:alg_general}. In particular, we can obtain an explicit solution for the density matrix at all times, as we show in the next subsection.

\subsection{Solution for the density matrix}

We now proceed with constructing a solution to the Lindblad equation of the form as in Eq.~(\ref{rho_anzats}). For this purpose it is convenient to exploit the structure of the algebra generated by superoperators (\ref{superops_def}).

First of all, one can easily check that the ${\mathfrak u}(1)$ generator is simply the linear Casimir invariant given by
\be \label{generator_u1}
	 Y_0 = X_1 - X_2 - X_3 + X_4,
\ee
which commutes with all other superoperators from Eq.~(\ref{superops_def}).
Second, for the $\mathfrak{sl}(4,{\mathbb C})$ subalgebra it is useful to choose a basis in the following way:
\be \label{superops_Y}
\begin{aligned}
 &Y_1=-X_{10},  && Y_6 = -X_6,  && Y_{11}=X_9, \\
 &Y_2=X_{13},  && Y_7=-(X_2 + X_4),  && Y_{12}=-X_{11}, \\
 &Y_3=-X_7,  && Y_8=  X_1 + X_2,  && Y_{13}=-X_5, \\
 &Y_4=-X_{14},  && Y_9=-(X_1 + X_3),  && Y_{14}=X_{12}, \\
 &Y_5=X_{16},  && Y_{10}=-X_8,  && Y_{15}=-X_{15}.\\
\end{aligned}
\ee
The above choice of the basis is motivated by the results of Ref.~\cite{Charzyski2013} and corresponds to the Cartan decomposintion of the algebra $\mathfrak{sl}(4, \mathbb{C})$ with respect to the Cartan subalgebra spanned by the generators $Y_7$, $Y_8$, and $Y_9$.
In terms of superoperators (\ref{generator_u1}) and (\ref{superops_Y}), the Liouvillian in Eq.~(\ref{Liouvillian_gens}) reads as:
\be \label{eq:Liouvillian_newbasis}
	{\cal L}_{k} = -\frac{\Lambda_k(t)}{2} {\mathds 1} + \sum_{j=0}^{15} a_j(t) Y_j,
\ee
where the coefficients $a_j$ are given by
\be \label{disentanglement_coeffs}
\begin{aligned}
   &a_0=\frac{i}{2} \left(\xi _k - \xi _{-k}\right), && a_1=-\mathfrak{c}_{-k},  \\
   &a_2 =\mathfrak{d}_{-k}, && a_3= \frac{1}{2} \left(\mathfrak{b}_k^*+4 \Delta _k\right),\\
   &a_4 = - a_2^*, && a_5 = \mathfrak{e}_{-k},\\
   &a_6 = a_3^*, && a_7 = \frac{1}{2} \left(\mathfrak{a}_{-k}+i \xi _{-k}+i \xi _k\right),\\
   &a_8 = \frac{1}{2} \left(\mathfrak{a}_{-k}-\mathfrak{a}_k\right), && a_9 = a_7^*,\\
   &a_{10} = \frac{1}{2} \left(\mathfrak{b}_k^*-4 \Delta _k\right),&& a_{11} = \mathfrak{c}_k,\\
   &a_{12} = -\mathfrak{d}_k^*, && a_{13} = a_{10}^*, \\
   &a_{14} = -a_{12}, && a_{15} = -\mathfrak{e}_k.
\end{aligned}
\ee
Let us emphasize once again that the method we are going to use allows one to deal with arbitrary time dependence of the parameters, and hence all $a_j$ in Eqs.~(\ref{eq:Liouvillian_newbasis}), (\ref{disentanglement_coeffs}) can be time-dependent.

We now employ the algorithm described in subsection~\ref{S:alg_general} and seek the solution to the Lindblad equation~(\ref{Lindblad_eq}) in the following form
\be \label{ansatz_rho_u1_sl4c}
\begin{aligned}
	\rho(t) &= \prod_{k\in \text{BZ}} \rho_k(t),\quad \text{where}\\
	\rho_k(t) &= e^{f(t)  {\mathds 1} + u_0(t) Y_0 } e^{u_{1}(t) Y_{1}} \ldots e^{u_{15}(t) Y_{15}} \rho_k(0)
\end{aligned}
\ee
and the functions $f(t)$ and $u_j(t)$ are yet unknown time- and momentum-dependent functions. Essentially, the Ansatz in Eq.~(\ref{ansatz_rho_u1_sl4c}) is nothing else than the Gauss parametrization of $SL(4, {\mathbb C})$. Differentiating Eq.~(\ref{ansatz_rho_u1_sl4c}) and taking into account that $\rho_k$ for different~$k$ commute, we obtain
\be \label{dot_rho}
	\dot\rho(t) = \sum_{k\in \text{BZ}} \dot \rho_k (t) \prod_{q\neq k} \rho_q (t).
\ee
Keeping in mind that $Y_0$ commutes with all other generators $Y_j$, we have for $\dot\rho_k$:
\be \label{dot_rho_k}
\begin{aligned}
	&\dot\rho_k = \Bigl( \dot f {\mathds 1} + \dot u_0 Y_0 + \dot u_1 Y_1 + \dot u_2 e^{u_1 \text{ad}_{Y_1}}Y_2  \,+  \\
	&+ \sum_{j=3}^{15} \dot u_j e^{u_1 \text{ad}_{Y_1}} \ldots e^{u_{j-1} \text{ad}_{Y_{j-1}}} Y_j \Bigr) \rho_k,
\end{aligned}
\ee
where, in order to obtain the last term on the first line and the second line, we used Eq. (\ref{Ad_action}). Then, using the results of Appendix~\ref{A:adjoint} to calculate various adjoint actions in the second line of Eq. (\ref{dot_rho_k}), we rewrite Eq. ~(\ref{dot_rho}) as
\be
	\dot \rho(t) = \sum_{k\in \text{BZ}} \tilde{\cal L}_k(t) \rho (t),
\ee
where we denoted
\be \label{tilde_L_k}
	\tilde{\cal L}_k(t) = \dot f(t) {\mathds 1} + \dot u_0(t) Y_0 + \sum_{j=1}^{15} \varphi_j\bigl(\dot u_j(t) , \{u_l(t)\}\bigr) Y_j.
\ee
The coefficients $\varphi_j\bigl(\dot u_j(t) , \{u_l(t)\}\bigr)$ result from the adjoint actions in Eq.~(\ref{dot_rho_k}). They are linear in $\dot u_j$, but can be nonlinear in $u_l(t)$.

For the density matrix in Eq.~(\ref{ansatz_rho_u1_sl4c}) to satisfy the Lindblad equation~(\ref{Lindblad_eq}) with the Liouvillian given by Eqs.~(\ref{Liouvillian_sum_k}) and (\ref{eq:Liouvillian_newbasis}), we have to require that the superoperator $\tilde {\cal L}_k$ in Eq. (\ref{tilde_L_k}) coincides [up to a factor of $1/2$, see Eq. (\ref{Liouvillian_sum_k})] with the Liouvillian~(\ref{eq:Liouvillian_newbasis}). Then, by matching the coefficients of ${\mathds 1}$ and $Y_j$, we obtain a set of equations for the functions $f(t)$ and $u_j(t)$:
\be \label{disentanglement_equations}
\begin{aligned}
	&\dot f(t) = -\Lambda_k(t)/4, \qquad \dot u_0 (t) = a_0(t)/2, \\
	&\varphi_j\bigl(\dot u_j(t) , \{u_l(t)\}\bigr) = a_j(t)/2 \qquad (j =1, \ldots, 15),
\end{aligned}
\ee
where the inhomogeneities $a_j$ are given by Eq.~(\ref{disentanglement_coeffs}). Obviously, all initial conditions are zero, $f(0) = u_j(0) = 0$.
For the sake of readability, the explicit form of the equations for $u_j$ in the second line of Eq.~(\ref{disentanglement_equations}) is presented in Appendix ~\ref{A:disentanglement_equations}. Here we only mention that the equations for $u_1$, $u_2$, and $u_3$ form a system of three coupled Riccati equations; Similarly, $u_4$ and $u_5$ satisfy a system of two coupled Riccati equations; Finally, $u_6$ obeys a scalar Ricatti equation, and all other functions $u_7, \ldots , u_{15}$ satisfy linear first order equations that can be directly integrated once the solutions for the preceding functions are known.

To summarize, we have reduced the Lindblad (operator) equation~(\ref{Lindblad_eq}) to a set of scalar equations~(\ref{disentanglement_equations}). Once the solutions to equations~(\ref{disentanglement_equations}) are known, one immediately obtains the density matrix at all times from Eq.~(\ref{ansatz_rho_u1_sl4c}).

A few comments are in order.
First of all, in general the coupled Riccati equations in Eq.~(\ref{disentanglement_equations}) can not be solved in quadratures, and one has to solve them numerically. Nevertheless,  it might be easier to gain insight into the physics and analyze the influence of various parameters via the solution of Eqs.~(\ref{disentanglement_equations}), rather than by a direct numerical integration of the (vectorized) Lindblad equation in some basis.

Second, in order to proceed further one should specify the initial condition, $\rho(0)$. Then, the action of the exponential factors in Eq.~(\ref{ansatz_rho_u1_sl4c}) can be calculated. However, this is beyond the scope of the present paper.

Finally, we would like to comment on the applicability and the advantage of the Lie-algebraic method that we have used. As we discussed in subsection~\ref{S:alg_general}, the method can be applied to any dissipative quantum system whose Liouvillian can be expressed as a linear combination of the generators of some closed Lie algebra. The method is especially useful in bosonic theories, where the underlying Hilbert space is infinitely dimensional, and numerical solution can only be obtained if one truncates the Hilbert space dimensionality.

For fermionic theories that we are dealing with, the Hilbert space is finite dimensional, and this allows one to construct the Liouvillian using a faithful matrix representation in a physically meaningful basis. We do so in the next Section, with the purpose of investigating the Liouvillian spectral properties.

\section{Liouvillian spectrum}\label{S:Liouvillian}

We now proceed with constructing a faithful matrix representation for the superoperators (\ref{superops_def}) in order to find the Liouvillian spectrum.
For a given $k>0$ the Hilbert space is four-dimensional and we choose a basis spanned by the vectors
\be \label{basis_vecs}
\begin{aligned}
	\ket{0_k, 0_{-k}} &= \left( \begin{array}{c} 0 \\ 1 \end{array} \right) \otimes \left( \begin{array}{c} 0 \\ 1 \end{array} \right) = (0, 0, 0, 1)^T,\\
	\ket{1_k, 0_{-k}} &= \left( \begin{array}{c} 1 \\ 0 \end{array} \right) \otimes \left( \begin{array}{c} 0 \\ 1 \end{array} \right) = (0, 1, 0, 0)^T,\\
	\ket{0_k, 1_{-k}} &= \left( \begin{array}{c} 0 \\ 1 \end{array} \right) \otimes \left( \begin{array}{c} 1 \\ 0 \end{array} \right) = (0, 0,1, 0)^T,\\
	\ket{1_k, 1_{-k}} &= \left( \begin{array}{c} 1 \\ 0 \end{array} \right) \otimes \left( \begin{array}{c} 1 \\ 0 \end{array} \right) = (1, 0, 0, 0)^T.\\
\end{aligned}
\ee

The fermionic annihilation and creation operators for the modes $k$ and $-k$ in this basis are represented by 
\be \label{fermionic_ops_matrix_rep}
\begin{aligned}
	c_{k} &= \sigma^{-} \otimes \sigma_0, \qquad c_{-k} = -\sigma_3 \otimes \sigma^{-},\\
	c^{\dag}_{k} &= \sigma^{+} \otimes \sigma_0, \qquad c^{\dag}_{-k} = -\sigma_3 \otimes \sigma^{+},
\end{aligned}
\ee
where $\otimes$ is the tensor product, $\sigma^{\pm} = (\sigma_1 \pm i \sigma_2)/2$, and $\sigma_{j}$ are the Pauli matrices. 
One can easily check that the matrices in Eq.~(\ref{fermionic_ops_matrix_rep}) satisfy the canonical anticommutation relations. The minus signs in the matrix representations for $c_{-k}$ and $c_{-k}^{\dag}$ in Eq. (\ref{fermionic_ops_matrix_rep}) are needed to preserve the signs in the mapping between the basis states in Eq. (\ref{basis_vecs}).
Then, using the well-known vectorization property
\be
	\vec{A B C} = \left( C{}^T \otimes A \right) \vec{B},
\ee
 we write
\be \label{X_j_vec}
	\vec{ X_j \rho } \equiv {\bf X}_j \vec{\rho},
\ee
where $\vec{\rho}$ is a vector formed from the columns of $\rho$ and the matrices ${\bf X}_j$ are given in Appendix \ref{A:matrix_rep}.
Then, applying the outlined vectorization procedure to the Lindblad equation, we obtain
\be
	\vec{\dot \rho} = \frac{1}{2} \sum_{ k \in \text{BZ} }  {\mathbb L}_k \vec{\rho},
\ee
where ${\mathbb L}_k$ is a matrix representation for the Liouvillian ${\cal L}_k$ from Eq.~(\ref{Liouvillian_gens}). Its explicit form is given in Appendix~\ref{A:matrix_rep}.

It is now straightforward to find the spectrum of ${\mathbb L}_k$ by direct diagonalization, which yields:
\be \label{spectrum}
\begin{aligned}
	&\lambda_0 = 0, \quad \lambda_{1,2} = -\frac{1}{2} \Lambda_k \pm i \left( \xi_k - \xi_{-k} \right),\\
	&\lambda_{3} = \lambda_1 + \lambda_2,\\
	&\lambda_{4,5} = -\frac{1}{4} \left( \Lambda_k + 2i \left( \xi_k - \xi_{-k} \right) \pm \sqrt{ U_k + i V_k} \right), \\
	&\lambda_{6,7} = -\frac{1}{4} \left( \Lambda_k - 2i \left( \xi_k - \xi_{-k} \right) \pm \sqrt{ U_k - i V_k} \right), \\
	&\lambda_{8,9} = -\frac{1}{4} \left( 3 \Lambda_k - 2i \left( \xi_k - \xi_{-k} \right) \pm \sqrt{ U_k + i V_k} \right), \\
	&\lambda_{10,11} = -\frac{1}{4} \left( 3 \Lambda_k + 2i \left( \xi_k - \xi_{-k} \right) \pm \sqrt{ U_k - i V_k} \right), \\
	&\lambda_{12,13} = -\frac{1}{4} \left( 2 \Lambda_k \pm \sqrt{2}\sqrt{ U_k +\sqrt{ U_k^2 + V_k^2}} \right),\\
	&\lambda_{14,15} = -\frac{1}{4} \left( 2 \Lambda_k \pm i \sqrt{2}\sqrt{- U_k + \sqrt{ U_k^2 + V_k^2}} \right),
\end{aligned}
\ee
where $\Lambda_k$ is given by Eq.~(\ref{Lambda_k}) and we denoted
\be \label{Uk_Vk}
\begin{aligned}
	U_k  &= \Theta_k^2 +  \left|\Phi_k \right|^2 -  \left[4 \left(\xi_k + \xi_{-k} \right)^2 + \left|8\Delta_k\right|^2 \right], \\
	V_k &=   2\Bigl[ 2\left( \xi_k + \xi_{-k}  \right)\Theta_k + \text{Im} \left\{ (8\Delta_k) ^* \Phi_k \right\} \Bigr],
\end{aligned}
\ee
with
\be \label{ThetaPhi}
\begin{aligned}
	&\Theta_k = |u_k|^2 - |v_k|^2 - |u_{-k}|^2  + |v_{-k}|^2, \\
	&\Phi_k = 2 \left( u_k v^*_k + u_{-k}v^*_{-k} \right), \\
\end{aligned}
\ee
and $\xi_k$, $\Delta_k$ given by Eq.~(\ref{xi_Delta_k}).

The functions $\Lambda_k$, $\Theta_k$, and $\Phi_k$ are related to each other via
\be \label{ThetaLambdaPhiPsi}
	\Theta_k^2 + \left| \Phi_k \right|^2 = \Lambda_k^2 - \left| \Psi_k \right|^2,
\ee
with
\be
	\Psi_k = 2\left( u_k u_{-k}^* - v_k v_{-k}^* \right),
\ee
and thus $U_k \leq \Lambda^2$. Taking into account that
\be \label{sqrt_ReIm}
\begin{aligned}
	\Re{ \sqrt{x + i y} } &= \frac{1}{\sqrt{2}} \sqrt{x+\sqrt{x^2 + y^2}}, \\
	\Im{ \sqrt{x + i y} } &= \frac{\text{sgn}(y)}{\sqrt{2}} \sqrt{-x+\sqrt{x^2 + y^2}},
\end{aligned}
\ee
one finds the following relations between the eigenvalues:
\be
\begin{aligned}
	\lambda_2 &= \lambda_1^*, && \lambda_{6,7} = \lambda_{4,5}^*, \\
	\lambda_{8,9} &= \lambda_1 + \lambda_{4,5}, && \lambda_{10,11} = \lambda_{8,9}^*, \\
	\lambda_{12,13} &= 2\text{Re}\{ \lambda_{4,5} \}, && \lambda_{14} = \lambda_{15}^* = \lambda_4 + \lambda_5^*.
\end{aligned}
\ee
Therefore, the whole spectrum can be determined from the knowledge of $\lambda_1$, $\lambda_4$, and $\lambda_5$. Moreover, the eigenvalues can be grouped into subbands by their real parts. Denoting $\nu_j = \Re{\lambda_j}/\Lambda_k$, we have:
\be
\begin{aligned}
	&-\frac{1}{2} \leq \nu_{4,6} \leq -\frac{1}{4}, \qquad -\frac{1}{4} \leq \nu_{5,7} \leq 0, \\
	&-1 \leq \nu_{8,10} \leq -\frac{3}{4}, \qquad  -\frac{3}{4} \leq \nu_{9,11} \leq -\frac{1}{2},\\
	&-1 \leq \nu_{12} \leq -\frac{1}{2}, \qquad  -\frac{1}{2} \leq \nu_{13} \leq 0.\\
\end{aligned}
\ee
Thus, we immediately see that if $\Lambda_k \neq 0$, the dissipative gap closes when $\Re{\lambda_{5,7}} = 0$, which also forces the real part of $\lambda_{13}$ to vanish. On the other hand, if for some $k$ one has $\Lambda_k =0$, then $\Re{\lambda_j} = 0$ for all $j$ \cite{proof}. 
In other words, if there exists a momentum mode at which the dissipative gap closes, then there is either a quadruple degeneracy, $\lambda_0 = \Re{\lambda_5} = \Re{\lambda_7} =\Re{\lambda_{13}} =0$, or a total degeneracy where at this momentum mode all eigenvalues have a vanishing real part.

\subsection{Closure of the dissipative gap}\label{S:gap}

Let us now analyze the dissipative gap and conditions of its closure in more detail. As can be seen from Eq.~(\ref{spectrum}), in the general case the gap closure condition reads as $\Lambda_k = \Re{ \sqrt{U_k \pm i V_k} }$, which yields
\be
	\Lambda_k^4 - U_k \Lambda_k^2 - V_k^2/4 = 0,
\ee
or, taking into account Eqs. (\ref{Uk_Vk}), (\ref{ThetaLambdaPhiPsi}), and (\ref{sqrt_ReIm}),
\be \label{gap_closure_generic}
	\Lambda_k^2 \left( \left| \Psi_k \right|^2 + 4 \left(\xi_k + \xi_{-k} \right)^2 + \left|8\Delta_k\right|^2 \right) = V_k^2 / 4.
\ee

In the case of purely dissipative evolution, i.e., $\xi_k = \Delta_k \equiv 0$, one has $V_k=0$ and Eq.~(\ref{gap_closure_generic}) reduces to $\Lambda_k \left| \Psi_k \right| = 0$. 
Then, the dissipative gap closes if $\Lambda_k = 0$, leading to the totally degenerate case, or if $u_k u_{-k}^* = v_k v_{-k}^*$, leading to the quadruple degeneracy.

In the simplest case, in which the coefficients $u_{j-n}$ and $v_{j-n}$ in Eq.~(\ref{Jump_op_real_space}) are real, condition~\eqref{gap_closure_generic} simplifies. 
The Fourier components then obviously satisfy
\be \label{real_u_v}
	u_{-k} =  u^*_k, \qquad v_{-k} =  v^*_k,
\ee
which leads to $\Theta_k = 0$ and $\Phi_k = 4 \Re{ u_k v_k^* }$.
Note that property (\ref{real_u_v}) is violated by the gauge transformation~(\ref{gauge_transformation}), and for this reason the case of complex $\Delta_k$ should be treated with care.
Assuming that $\Delta_k$ is real, we have $V_k = 0$ and
the gap closure condition (\ref{gap_closure_generic}) can be satisfied by
\be \label{L_real_gap_closure}
	\xi_k + \xi_{-k} = \Delta_k = 0\quad \text{ and } \quad  u_k = v_k.
\ee

Below we consider more non-trivial cases of the gap closure.

\subsection{Gapless modes with a linear spectrum}

Let us consider fermions with zero pairing ($\Delta_k = 0$) and a long-range hopping of the following form:
\be \label{long_range_hopping}
	\xi_k = \mu +
	\sum_{n=1}^{+\infty}\frac{2 \cos k n}{n^{\alpha}} =\mu + 2 \text{Ci}_{\alpha}k,
\ee
where $\text{Ci}_{\alpha}k$ is the Clausen function, which is related to the polylogarithm as $2\text{Ci}_{\alpha}k=  \text{Li}_{\alpha}e^{i k} + \text{Li}_{\alpha}e^{-i k}$. For $1< \alpha < 3$ it has the following expansion for small $k$:
\be \label{Clausen_Taylor}
	\text{Ci}_{\alpha}k \approx \zeta(\alpha) +  \Gamma(1-\alpha) \sin\frac{\pi \alpha}{2} \left| k \right|^{\alpha-1}-\zeta(\alpha-2)\frac{k^2}{2},
\ee
where $\zeta(x)$ is the Riemann $\zeta$-function.

For simplicity we take a jump operator of the form
\be \label{jump_simple}
	L_j = \sqrt{g} \left( c_j + c_j^{\dag} \right),
\ee
such that $u_k = v_k = \sqrt{g}$. 
This corresponds to the case described above.  
We have  $\Lambda_k = \Phi_k = 4 g$, $\Theta_k = 0$. Therefore $V_k =0$  and $U_k = 16 (g^2 - \xi_k^2)$. For $U_k \geq 0$ we thus have:
\be
	\Delta_{\text{d}} \equiv \Re{ \lambda_{5,7} }= \frac{1}{2} \Re{\lambda_{13}} = - g + \sqrt{ g^2 - \xi_k^2}.
\ee
\begin{figure}[t]
\includegraphics[width=8cm]{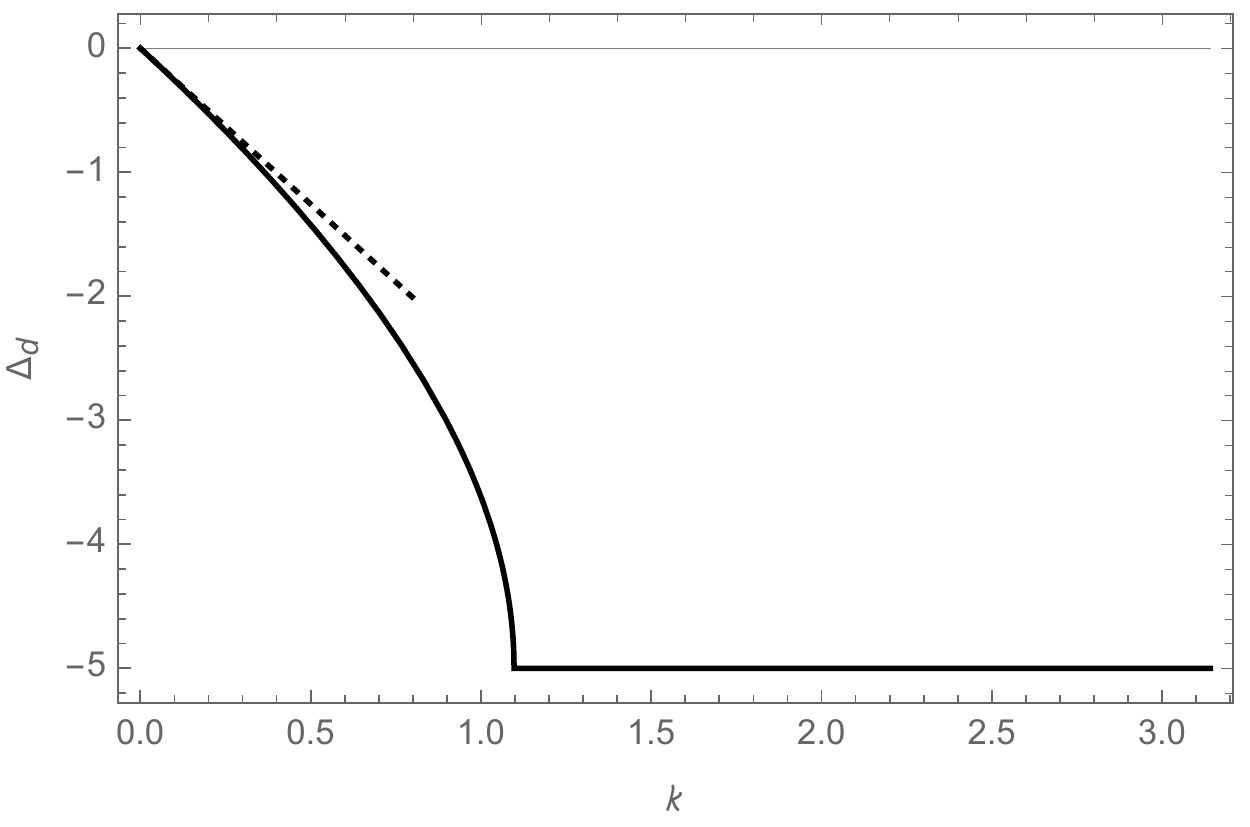}
\caption{The dissipative gap $\Delta_{\text{d}}$ versus momentum $k$ for the long-range hopping model [see Eq.~(\ref{long_range_hopping})] with the dissipation described by jump operator~(\ref{jump_simple}). The dotted line shows the asymptotic linear behaviour according to Eq.~(\ref{lin_gap}). The dissipation strength is $g=5$.}
\label{F:linear}
\end{figure}
The dissipative gap $\Delta_{\text{d}}$ closes if $\xi_k = 0$. From Eqs.~(\ref{long_range_hopping}) and (\ref{Clausen_Taylor}) we immediately see that this happens at $k = 0$ for $\mu = -2\zeta(\alpha)$, and sufficiently close to zero the gap behaves as $|k|^{2(\alpha-1)}$, which for $\alpha = 3/2$ gives
\be \label{lin_gap}
	\Delta_{\text{d}} \approx -\frac{4\pi}{g} |k|, \qquad \mu = -2 \zeta(3/2).
\ee
This linear behaviour is demonstrated in Fig.~\ref{F:linear}.
For $\mu < -2 \zeta(3/2)$ the gap is open ($\Delta_{\text{d}} \neq 0$), whereas as one increases the chemical potential beyond $\mu=-2\zeta(3/2)$, the gap closure point at $k=0$ splits in two symmetrical points at $\pm k_*$, satisfying $\text{Ci}_{3/2}k_* = -\mu/2$. In the vicinity of these points the gap behaviour changes from linear to quadratic:
\be
	\Delta_{\text{d}} \approx -\frac{1}{2g} \left[ i \text{Li}_{1/2}e^{i k_*} - i \text{Li}_{1/2}e^{-i k_*} \right]^2 (|k|-k_*)^2.
\ee
 The function in the square brackets is non-negative for all $0 < k_* \leq \pi$, and it vanishes only for $k_* = \pi$, corresponding to $\mu = (2-\sqrt{2})\zeta(3/2)$. Thus, near the boundaries of the Brillouin zone the gap behaves as
 \be
 	\Delta_{\text{d}} \approx -\frac{\left(9 - 4 \sqrt{2}\right)  \zeta \left(-\frac{1}{2}\right)^2}{2 g} (|k|-\pi )^4.
 \ee
 Finally, for $\mu > (2-\sqrt{2})\zeta(3/2)$ the gap opens again.
 
 It would be interesting to investigate whether such regimes can be related to non-unitary conformal field theories. However, this question is beyond the scope of the present work and we leave it to future investigations.
 
 \subsection{Multiple gap closure points}
 
An interesting situation appears when the spectrum closes simultaneously at zero and finite momenta. 
In these cases the Liouvillian spectrum exhibits a feature resembling rotons, which are a special kind of elementary excitation forming a minimum of energy at finite momentum in quantum liquids, such as $^4$He and dipolar quantum ensembles.

This ``roton-like" form of the Liouvillian spectrum can be achieved in a number of different ways. Here we restrict ourselves to one of the simplest cases, namely, the dissipative Kitaev model described by Hamiltonian~(\ref{H_k}) with 
\be
	\xi_k = \mu + 2 \cos k, \quad \Delta_k = \gamma \sin k
\ee
and the following jump operator

\be \label{Jump_op_roton}
	L_j = \sqrt{g} \left[ c_j + \delta_1 c_j^{\dag} + \delta_2 \left( c_{j+1} + \delta_3 c_{j+1}^{\dag} \right) \right].
\ee

\begin{figure}[t]
\includegraphics[width=8cm]{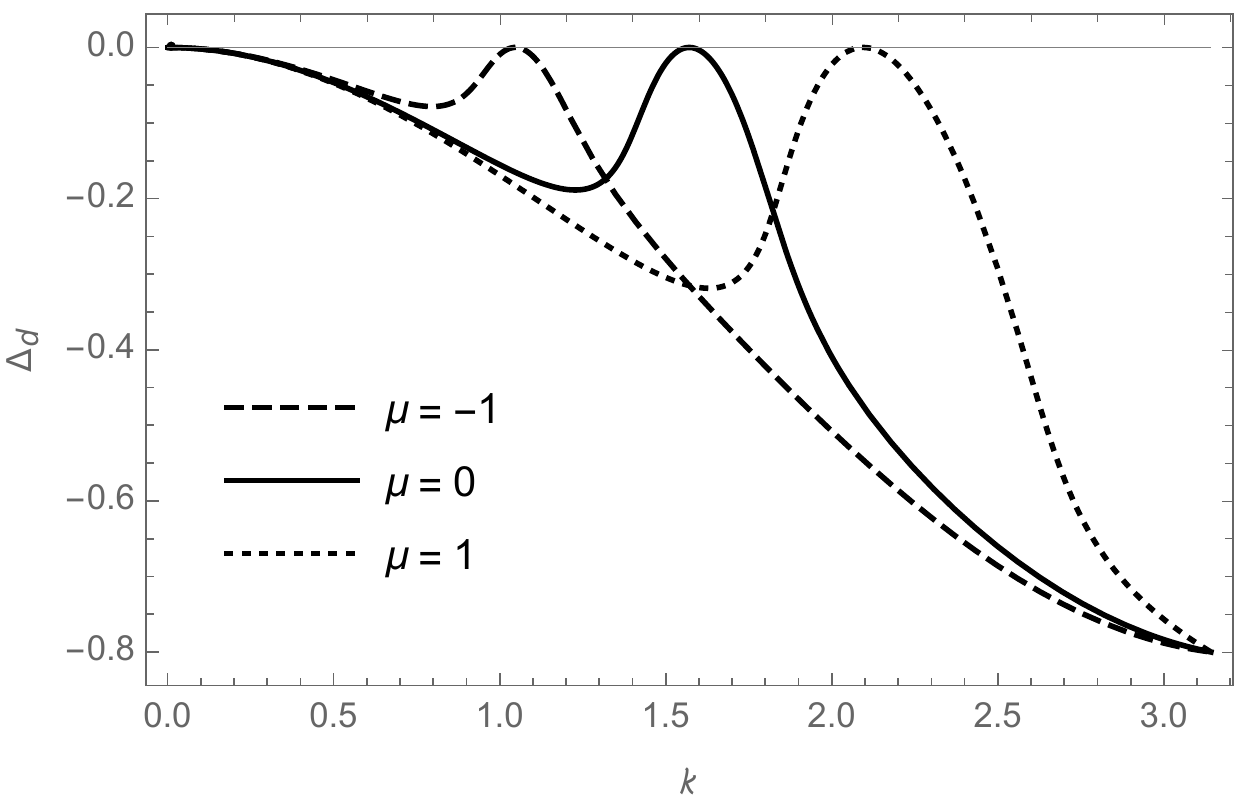}
\caption{The dissipative gap $\Delta_{\text{d}}$ versus momentum $k$ for the Kitaev model with the dissipation described by jump operator~(\ref{Jump_op_roton}). One can clearly see the ``Roton-like" feature leading to the simultaneous gap closure at zero and finite momenta. The choice of parameters corresponds to Eq.~(\ref{roton_params}). The pairing amplitude is $\gamma = 0.2$ and the dissipation strength is $g=0.1$.}
\label{F:roton}
\end{figure}
In the momentum space [see Eq. (\ref{Jump_op_k_space})] it has the coefficients 
\be
	u_k = \sqrt{g} \left( 1 + \delta_2 e^{i k} \right), \quad v_k = \sqrt{g} \left( \delta_1 + \delta_2 \delta_3 e^{i k} \right).
\ee
Using Eqs. (\ref{spectrum}) one can easily show that by choosing
\be \label{roton_params}
	\delta_1 = \delta_3 = i, \qquad \delta_2 = -1, 
\ee
for an arbitrary value of the pairing amplitude $\gamma$ and $-2 \leq \mu \leq 2$
the dissipative gap closes simultaneously at $k=0$ and at a non-zero momentum inside the Brillouin zone. This is demonstrated in Fig.~\ref{F:roton}. In the vicinity of both gap closure points the gap behaves as $\Delta_{\text{d}} \sim k^2$. 
%

The configurations discussed in this subsection are promising for the realization of quantum computing algorithms in the presence of noise. 

\section{Conclusion and outlook}\label{S:Conclusion}

In conclusion, we have investigated dissipative translationally-invariant free fermionic theories with quadratic Liouvillians.
We have demonstrated the applicability of the Lie-algebraic approach for the description of dissipative translationally-invariant free fermionic theories with quadratic Liouvillians. 
We have derived the criterion for the closure of the dissipative gap, which is believed to be linked with nonequilibrium dissipative phase transitions. 
We have also provided a few examples of exotic regimes of the spectrum closure: (i) the presence of gapless modes with a linear spectrum for fermions with a long-range hopping, which might be related to non-unitary conformal field theories;
(ii) non-monotonic roton-like spectrum closure, which is promising for the realization of quantum computing algorithms in the presence of noise. 

Further directions of our studies are related to including the consideration of topological effects in the consideration. 
In addition, it is an interesting point to understand the potential role of the obtained configurations with the non-monotonic spectrum closure for quantum computing.

We expect that the predicted effects can be probed in experiments with ultracold atomic and quantum-optical systems using currently available experimental facilities. 
Recently proposed setups for the realization of the Kitaev model using systems with a sufficient degree of tunability, such as atomic quantum wires and arrays of identical nonlinear cavities coupled through nearest-neighbor photon tunneling.
These setups can be extended for the realization of the model, which is considered in our work, which makes it realistic to observe predicted phenomena. 

\section*{Acknowledgments} 
We thank E. Vernier for drawing our attention to Ref.~\cite{Vernier2020}.
This work is part of the DeltaITP consortium, a program of the Netherlands Organization for Scientific Research (NWO) that is funded by
the Dutch Ministry of Education, Culture and Science
(OCW).
The results of D.V.K. and V.I.Y. on the application of the Lie-algebraic approach and studying the spectrum properties were supported by the Russian Science Foundation Grant No. 19-71-10091 
(parts of Sec.~\ref{S:dynamical_symmetry} and Sec.~\ref{S:Liouvillian}). 
The work of A.K.F. was supported by grant UMNIK (Agreement 103GUCEC8-D3/56361 form 21.12.2019).


\appendix
\begin{widetext}
\section{Superoperator algebra} \label{A:algebra}

In this Appendix we present a table with the commutation relations for the algebra generated by the superoperators defined in Eq.~(\ref{superops_def}).
\begin{align*}
    \begin{array}{c|cccccccccccccccc}
   & X_{1} & X_{2} & X_{3} & X_{4} & X_{5} & X_{6} & X_{7} & X_{8} & X_{9} & X_{10} & X_{11} & X_{12} &
   X_{13} & X_{14} & X_{15} & X_{16} \\\hline
 X_{1} & 0 &  &  &  &  &  &  &  &  &  &  &  &  &  &  &  \\
 X_{2} & 0 & 0 &  &  &  &  &  &  & & &  &  &  &  &  &  \\
 X_{3} & 0 & 0 & 0 &  &  &  &  &  &  &  &  &  &  &  &  &  \\
 X_{4} & 0 & 0 & 0 & 0 &  &  &  &  &  &  &  &  &  &  &  &  \\
 X_{5} & -X_{5} & 0 & -X_{5} & 0 & 0 &  &  &  &  &  &  &  &  &  &  & 
   \\
 X_{6} & 0 & X_{6} & 0 & X_{6} & 0 & 0 &  &  & &  &  &  &  &  &  & 
   \\
 X_{7} & X_{7} & 0 & X_{7} & 0 & -A & 0 & 0 &  &  &  &  &  &  &  &  & 
   \\
 X_{8} & 0 & -X_{8} & 0 & -X_{8} & 0 & B & 0 & 0 &  &  &  &  &  &  &  & 
   \\
 X_{9} & X_{9} & X_{9} & 0 & 0 & X_{12} & 0 & 0 & X_{11} & 0 &  &  &  &  &  &  &  \\
 X_{10} & 0 & 0 & X_{10} & X_{10} & -X_{14} & 0 & 0 & -X_{13} & 0 & 0 &  & &  &  &  & 
   \\
 X_{11} & X_{11} & 0 & 0 & -X_{11} & X_{15} & X_{9} & 0 & 0 & 0 & -X_{7} & 0 &  &  &  &  & 
   \\
 X_{12} & 0 & X_{12} & -X_{12} & 0 & 0 & 0 & X_{9} & X_{15} & 0 & -X_{6} & 0 & 0 & &  &  & 
   \\
 X_{13} & 0 & -X_{13} & X_{13} & 0 & -X_{16} & -X_{10} & 0 & 0 & X_{7} & 0 & 0 & F & 0 &  &  & 
   \\
 X_{14} & -X_{14} & 0 & 0 & X_{14} & 0 & 0 & -X_{10} & -X_{16} & X_{6} & 0 & E & 0 & 0 & 0 &  & 
   \\
 X_{15} & 0 & 0 & -X_{15} & -X_{15} & 0 & X_{12} & X_{11} & 0 & 0 & D & 0 & 0 & -X_{8} & -X_{5} & 0 & \\
 X_{16} & -X_{16} & -X_{16} & 0 & 0 & 0 & -X_{14} & -X_{13} & 0 & C & 0 & X_{8} & X_{5} & 0 & 0 & 0 & 0 \\
\end{array}
\end{align*}
Where we denoted
\begin{align*}
\begin{array}{lll}
    A = X_{1} + X_{3}, \hspace{2 cm} & C = X_{1}+X_{2}, \hspace{2 cm} &   E = X_{1}-X_{4},\\
    B = X_{2}+X_{4}, &   D = X_{3}+X_{4}, &   F = X_{2}-X_{3}.
\end{array}
\end{align*}

\section{Matrix representation} \label{A:matrix_rep}

In this Appendix we present a matrix representation for the superoperators defined in Eq.~(\ref{superops_def}) and for the Liouvillian in Eq. (\ref{Liouvillian_gens}). As discussed in Section~\ref{S:Liouvillian} of the main text, this matrix representation corresponds to choosing a basis~(\ref{basis_vecs}) in the Hilbert space for a given momentum mode $k$. Then, using Eqs.~(\ref{superops_def}) and (\ref{fermionic_ops_matrix_rep})--(\ref{X_j_vec}) one obtains
\be \label{X_mat_rep}
\begin{aligned}
   {\bf X}_1 &=  \sigma _0\otimes \sigma _0 \otimes \left(\sigma^{+}
   \sigma^{-}\right) \otimes \sigma _0 - \frac{1}{2} {\mathds 1},\qquad &&
   {\bf X}_2  =  \left(\sigma^{+} \sigma^{-}\right)\otimes \sigma _0\otimes \sigma
   _0\otimes \sigma _0 - \frac{1}{2} {\mathds 1}, \\
  {\bf X}_3 &=  \sigma _0\otimes \sigma _0\otimes \sigma _0\otimes
   \left(\sigma^{+} \sigma^{-}\right)- \frac{1}{2} {\mathds 1},  &&
 {\bf X}_4  = \sigma _0\otimes \left(\sigma^{+}\sigma^{-}\right)\otimes \sigma
   _0\otimes \sigma _0 - \frac{1}{2} {\mathds 1}, \\
  {\bf X}_5 &=	 -  \sigma _0\otimes \sigma _0 \otimes \left(\sigma^{+} \sigma
   _3\right) \otimes \sigma^{+}, &&
 {\bf  X}_6 =     -  \left(\sigma _3 \sigma^{-}\right) \otimes \sigma^{-} \otimes \sigma _0 \otimes
   \sigma _0, \\
  {\bf X}_7 &=  \, \sigma _0 \otimes \sigma _0 \otimes \left(\sigma^{-} \sigma
   _3\right)\otimes \sigma^{-}, &&
  {\bf X}_8 =   \, \left( \sigma _3  \sigma^{+} \right) \otimes \sigma^{+} \otimes \sigma _0 \otimes
   \sigma _0, \\
  {\bf X}_9 &=   \sigma^{-} \otimes \sigma _0 \otimes \sigma^{-} \otimes \sigma _0, &&
  {\bf X}_{10} =  \sigma _3 \otimes \sigma^{-} \otimes \sigma _3 \otimes \sigma^{-},  \\
  {\bf X}_{11}  &=  - \sigma _3 \otimes \sigma^{+} \otimes \sigma^{-} \otimes \sigma _0, && 
  {\bf X}_{12}  =  - \sigma^{-} \otimes \sigma _0 \otimes \sigma _3 \otimes \sigma^{+}, \\
  {\bf X}_{13}  &=  - \sigma^{+} \otimes \sigma _0\otimes \sigma _3\otimes \sigma^{-}, &&
  {\bf X}_{14}  =  -  \sigma _3 \otimes \sigma^{-} \otimes \sigma^{+} \otimes \sigma _0, \\
  {\bf X}_{15}  &=  \sigma _3 \otimes \sigma^{+}\otimes \sigma _3\otimes \sigma^{+}, &&
  {\bf X}_{16}  =  \sigma^{+} \otimes \sigma _0\otimes \sigma^{+} \otimes \sigma _0, \\
\end{aligned}
\ee
where $\otimes$ is the tensor product, $\sigma_{j}$ are the Pauli matrices, $\sigma^{\pm} = (\sigma_1 \pm i \sigma_2)/2$, and ${\mathds 1}$ is the $16\times16$ identity matrix.

Using Eq. (\ref{X_mat_rep}) and Eq. (\ref{Liouvillian_gens}) of the main text, we immediately obtain the following matrix representation for the Liouvillian itself:

\be
\begin{aligned}
{\mathbb L}_k =
\left(
\begin{array}{cccccccccccccccc}
 \mathfrak{D}_{k,1} & 0 & 0 & \mathfrak{A}_{k,+}^* & 0 & \mathfrak{e}_k & 0 & 0 & 0 & 0 &
   \mathfrak{e}_{-k} & 0 & \mathfrak{A}_{k,+} & 0 & 0 & 0 \\
 0 & \mathfrak{C}_k & 0 & 0 & 0 & 0 & 0 & 0 & -\mathfrak{d}_{-k} & 0 & 0 & \mathfrak{e}_{-k} & 0 &
   \mathfrak{A}_{k,+} & 0 & 0 \\
 0 & 0 & \mathfrak{C}_{-k} & 0 & -\mathfrak{d}_k & 0 & 0 & -\mathfrak{e}_k & 0 & 0 & 0 & 0 & 0 & 0 &
   \mathfrak{A}_{k,+} & 0 \\
 \mathfrak{A}_{k,-} & 0 & 0 & \mathfrak{F}_{k,+} & 0 & -\mathfrak{d}_k & 0 & 0 & 0 & 0 & \mathfrak{d}_{-k} &
   0 & 0 & 0 & 0 & \mathfrak{A}_{k,+} \\
 0 & 0 & -\mathfrak{d}_{-k}^* & 0 & \mathfrak{C}_k^* & 0 & 0 & \mathfrak{A}_{k,+}^* & 0 & 0 & 0 &
   0 & 0 & 0 & \mathfrak{e}_{-k} & 0 \\
 \mathfrak{c}_{-k} & 0 & 0 & -\mathfrak{d}_{-k}^* & 0 & \mathfrak{D}_{k,2} & 0 & 0 & 0 & 0 & 0 & 0 &
   -\mathfrak{d}_{-k} & 0 & 0 & \mathfrak{e}_{-k} \\
 0 & 0 & 0 & 0 & 0 & 0 & \mathfrak{F}_{k,-} & 0 & 0 & 0 & 0 & 0 & 0 & 0 & 0 & 0 \\
 0 & 0 & -\mathfrak{c}_{-k} & 0 & \mathfrak{A}_{k,-} & 0 & 0 & \mathfrak{E}_k & 0 & 0 & 0 & 0 & 0 & 0 &
   \mathfrak{d}_{-k} & 0 \\
 0 & -\mathfrak{d}_k^* & 0 & 0 & 0 & 0 & 0 & 0 & \mathfrak{C}_{-k}^* & 0 & 0 &
   \mathfrak{A}_{k,+}^* & 0 & -\mathfrak{e}_k & 0 & 0 \\
 0 & 0 & 0 & 0 & 0 & 0 & 0 & 0 & 0 & \mathfrak{F}_{k,-}^* & 0 & 0 & 0 & 0 & 0 & 0 \\
 \mathfrak{c}_k & 0 & 0 & \mathfrak{d}_k^* & 0 & 0 & 0 & 0 & 0 & 0 & \mathfrak{D}_{-k,2} & 0 &
   \mathfrak{d}_k & 0 & 0 & \mathfrak{e}_k \\
 0 & \mathfrak{c}_k & 0 & 0 & 0 & 0 & 0 & 0 & \mathfrak{A}_{k,-} & 0 & 0 & \mathfrak{E}_{-k} & 0 &
   \mathfrak{d}_k & 0 & 0 \\
 \mathfrak{A}_{k,-}^* & 0 & 0 & 0 & 0 & -\mathfrak{d}_k^* & 0 & 0 & 0 & 0 & \mathfrak{d}_{-k}^* &
   0 & \mathfrak{F}_{k,+}^* & 0 & 0 & \mathfrak{A}_{k,+}^* \\
 0 & \mathfrak{A}_{k,-}^* & 0 & 0 & 0 & 0 & 0 & 0 & -\mathfrak{c}_{-k} & 0 & 0 & \mathfrak{d}_{-k}^*
   & 0 & \mathfrak{E}_k^* & 0 & 0 \\
 0 & 0 & \mathfrak{A}_{k,-}^* & 0 & \mathfrak{c}_k & 0 & 0 & \mathfrak{d}_k{}^* & 0 & 0 & 0 & 0 & 0 &
   0 & \mathfrak{E}_{-k}^* & 0 \\
 0 & 0 & 0 & \mathfrak{A}_{k,-}^* & 0 & \mathfrak{c}_k & 0 & 0 & 0 & 0 & \mathfrak{c}_{-k} & 0 &
   \mathfrak{A}_{k,-} & 0 & 0 & \mathfrak{D}_{k,3} \\
\end{array}
\right),
\end{aligned}
\ee
where the functions $\mathfrak{c}_k, \mathfrak{d}_k$, and $\mathfrak{e}_k$ are given by Eq.~(\ref{aux_functions_abcde}) of the main text, and for brevity we introduced the following quantities:

\be
\begin{aligned}
\mathfrak{A}_{k,\sigma} &=\frac{\mathfrak{d}_{-k}}{2}-\frac{\mathfrak{d}_k}{2}-2\sigma \Delta _k & \mathfrak{D}_{k,1}&=-\mathfrak{c}_{-k}-\mathfrak{c}_k,\\
\mathfrak{C}_k&=-\frac{\mathfrak{c}_{-k}}{2}-\mathfrak{c}_k-\frac{\mathfrak{e}_k}{2}-i \xi _{-k}, & \mathfrak{D}_{k,2}&=-\mathfrak{c}_k-\mathfrak{e}_k,\\
\mathfrak{E}_k&=-\frac{\mathfrak{c}_k}{2}-\frac{\mathfrak{e}_{-k}}{2}-\mathfrak{e}_k-i \xi _k, & \mathfrak{D}_{k,3}&=-\mathfrak{e}_{-k}-\mathfrak{e}_k,\\
\mathfrak{F}_\sigma &=-\frac{\mathfrak{c}_{-k}}{2}-\frac{\mathfrak{c}_k}{2}-\frac{\mathfrak{e}_{-k}}{2}-\frac{\mathfrak{e}_k}{2}-i\sigma \xi _{-k}-i \xi _k,
\end{aligned}
\ee
with $\sigma = \pm 1$.
%

\end{widetext}

\section{Nontrivial adjoint actions for $\mathfrak{sl}(4,\mathbb{C})$} \label{A:adjoint}

In this Appendix we present all non-trivial adjoint actions of the superoperators defined in Eq.~(\ref{superops_Y}):

\begin{flalign*}
&\begin{array}{l}
 e^{\tau  \text{ad}_{Y_{1}}} Y_7 = Y_7 -\tau  Y_1 \\
 e^{\tau  \text{ad}_{Y_{1}}} Y_9 = Y_9 - \tau  Y_1 \\
 e^{\tau  \text{ad}_{Y_{1}}} Y_{10} = Y_{10} - \tau  Y_2 \\
 e^{\tau  \text{ad}_{Y_{1}}} Y_{12} = Y_{12} - \tau  Y_3 \\
 e^{\tau  \text{ad}_{Y_{1}}} Y_{13} = Y_{13} + \tau  Y_4 \\
 e^{\tau  \text{ad}_{Y_{1}}} Y_{14} = Y_{14} + \tau  Y_6 \\
 e^{\tau  \text{ad}_{Y_{1}}} Y_{15} = Y_{15} + \tau  \left(Y_7+Y_8+Y_9\right)  - \tau ^2 Y_1\\
\end{array}&
\end{flalign*}

\begin{flalign*}&\begin{array}{l}
 e^{\tau  \text{ad}_{Y_{2}}} Y_6= Y_6 - \tau  Y_1 \\
 e^{\tau  \text{ad}_{Y_{2}}} Y_7= Y_7 + \tau  Y_2 \\
 e^{\tau  \text{ad}_{Y_{2}}} Y_8= Y_8 - \tau  Y_2 \\
 e^{\tau  \text{ad}_{Y_{2}}} Y_9= Y_9 - \tau  Y_2 \\
 e^{\tau  \text{ad}_{Y_{2}}} Y_{11}= Y_{11} - \tau  Y_3 \\
 e^{\tau  \text{ad}_{Y_{2}}} Y_{13}=Y_{13} + \tau  Y_5 \\
 e^{\tau  \text{ad}_{Y_{2}}} Y_{14}= Y_{14} +\tau  \left(Y_8+Y_9\right) - \tau ^2 Y_2 \\
 e^{\tau  \text{ad}_{Y_{2}}} Y_{15}= Y_{15} + \tau  Y_{10} \\
\end{array}&\nonumber
\end{flalign*}

\begin{flalign*}
&\begin{array}{l}
 e^{\tau  \text{ad}_{Y_{3}}} Y_4= Y_4 - \tau  Y_1 \\
 e^{\tau  \text{ad}_{Y_{3}}} Y_5=Y_5 - \tau  Y_2 \\
 e^{\tau  \text{ad}_{Y_{3}}} Y_8= Y_8 + \tau  Y_3 \\
 e^{\tau  \text{ad}_{Y_{3}}} Y_9= Y_9 - 2 \tau  Y_3 \\
 e^{\tau  \text{ad}_{Y_{3}}} Y_{13}= Y_{13} +\tau  Y_9 - \tau ^2 Y_3 \\
 e^{\tau  \text{ad}_{Y_{3}}} Y_{14}= Y_{14} + \tau  Y_{11} \\
 e^{\tau  \text{ad}_{Y_{3}}} Y_{15}=  Y_{15}+ \tau  Y_{12} \\
\end{array}&
\end{flalign*}

\begin{flalign*}&\begin{array}{l}
 e^{\tau  \text{ad}_{Y_{4}}} Y_3= Y_3 + \tau  Y_1 \\
 e^{\tau  \text{ad}_{Y_{4}}} Y_7=  Y_7 - \tau  Y_4 \\
 e^{\tau  \text{ad}_{Y_{4}}} Y_8= Y_8 - \tau  Y_4 \\
 e^{\tau  \text{ad}_{Y_{4}}} Y_9= Y_9 + \tau  Y_4 \\
 e^{\tau  \text{ad}_{Y_{4}}} Y_{10}=  Y_{10} - \tau  Y_5 \\
 e^{\tau  \text{ad}_{Y_{4}}} Y_{11}=  Y_{11} + \tau  Y_6 \\
 e^{\tau  \text{ad}_{Y_{4}}} Y_{12}= Y_{12} +\tau  \left(Y_7+Y_8\right) - \tau ^2 Y_4 \\
 e^{\tau  \text{ad}_{Y_{4}}} Y_{15}= Y_{15} - \tau  Y_{13} \\
\end{array}&
\end{flalign*}

\begin{flalign*}
&\begin{array}{l}
 e^{\tau  \text{ad}_{Y_{5}}} Y_3= Y_3 + \tau  Y_2 \\
 e^{\tau  \text{ad}_{Y_{5}}} Y_6= Y_6 - \tau  Y_4 \\
 e^{\tau  \text{ad}_{Y_{5}}} Y_7= Y_7 + \tau  Y_5 \\
 e^{\tau  \text{ad}_{Y_{5}}} Y_8= Y_8 - 2 \tau  Y_5 \\
 e^{\tau  \text{ad}_{Y_{5}}} Y_9= Y_9 + \tau  Y_5 \\
 e^{\tau  \text{ad}_{Y_{5}}} Y_{11}= Y_{11} +\tau  Y_8 - \tau ^2 Y_5 \\
 e^{\tau  \text{ad}_{Y_{5}}} Y_{12}= Y_{12} + \tau  Y_{10} \\
 e^{\tau  \text{ad}_{Y_{5}}} Y_{14}= Y_{14} - \tau  Y_{13} \\
\end{array} &
\end{flalign*}

\begin{flalign*} &\begin{array}{l}
 e^{\tau  \text{ad}_{Y_{6}}} Y_2= Y_2 + \tau  Y_1 \\
 e^{\tau  \text{ad}_{Y_{6}}} Y_5= Y_5 + \tau  Y_4 \\
 e^{\tau  \text{ad}_{Y_{6}}} Y_7= Y_7 - 2 \tau  Y_6 \\
 e^{\tau  \text{ad}_{Y_{6}}} Y_8= Y_8 + \tau  Y_6 \\
 e^{\tau  \text{ad}_{Y_{6}}} Y_{10}= Y_{10} +\tau  Y_7 - \tau ^2 Y_6 \\
 e^{\tau  \text{ad}_{Y_{6}}} Y_{12}= Y_{12} - \tau  Y_{11} \\
 e^{\tau  \text{ad}_{Y_{6}}} Y_{15}= Y_{15} - \tau  Y_{14} \\
\end{array}&
\end{flalign*}

\begin{flalign*}
&\begin{array}{l}
 e^{\tau  \text{ad}_{Y_{7}}} Y_1=e^{\tau } Y_1 \\
 e^{\tau  \text{ad}_{Y_{7}}} Y_2=e^{-\tau } Y_2 \\
 e^{\tau  \text{ad}_{Y_{7}}} Y_4=e^{\tau } Y_4 \\
 e^{\tau  \text{ad}_{Y_{7}}} Y_5=e^{-\tau } Y_5 \\
 e^{\tau  \text{ad}_{Y_{7}}} Y_6=e^{2 \tau } Y_6 \\
 e^{\tau  \text{ad}_{Y_{7}}} Y_{10}=e^{-2 \tau } Y_{10} \\
 e^{\tau  \text{ad}_{Y_{7}}} Y_{11}=e^{\tau } Y_{11} \\
 e^{\tau  \text{ad}_{Y_{7}}} Y_{12}=e^{-\tau } Y_{12} \\
 e^{\tau  \text{ad}_{Y_{7}}} Y_{14}=e^{\tau } Y_{14} \\
 e^{\tau  \text{ad}_{Y_{7}}} Y_{15}=e^{-\tau } Y_{15} \\
\end{array} &
\end{flalign*}

\begin{flalign*}&\begin{array}{l}
 e^{\tau  \text{ad}_{Y_{8}}} Y_2=e^{\tau } Y_2 \\
 e^{\tau  \text{ad}_{Y_{8}}} Y_3=e^{-\tau } Y_3 \\
 e^{\tau  \text{ad}_{Y_{8}}} Y_4=e^{\tau } Y_4 \\
 e^{\tau  \text{ad}_{Y_{8}}} Y_5=e^{2 \tau } Y_5 \\
 e^{\tau  \text{ad}_{Y_{8}}} Y_6=e^{-\tau } Y_6 \\
 e^{\tau  \text{ad}_{Y_{8}}} Y_{10}=e^{\tau } Y_{10} \\
 e^{\tau  \text{ad}_{Y_{8}}} Y_{11}=e^{-2 \tau } Y_{11} \\
 e^{\tau  \text{ad}_{Y_{8}}} Y_{12}=e^{-\tau } Y_{12} \\
 e^{\tau  \text{ad}_{Y_{8}}} Y_{13}=e^{\tau } Y_{13} \\
 e^{\tau  \text{ad}_{Y_{8}}} Y_{14}=e^{-\tau } Y_{14} \\
\end{array}&
\end{flalign*}

\begin{flalign*}
&\begin{array}{l}
 e^{\tau  \text{ad}_{Y_{9}}} Y_1=e^{\tau } Y_1 \\
 e^{\tau  \text{ad}_{Y_{9}}} Y_2=e^{\tau } Y_2 \\
 e^{\tau  \text{ad}_{Y_{9}}} Y_3=e^{2 \tau } Y_3 \\
 e^{\tau  \text{ad}_{Y_{9}}} Y_4=e^{-\tau } Y_4 \\
 e^{\tau  \text{ad}_{Y_{9}}} Y_5=e^{-\tau } Y_5 \\
 e^{\tau  \text{ad}_{Y_{9}}} Y_{11}=e^{\tau } Y_{11} \\
 e^{\tau  \text{ad}_{Y_{9}}} Y_{12}=e^{\tau } Y_{12} \\
 e^{\tau  \text{ad}_{Y_{9}}} Y_{13}=e^{-2 \tau } Y_{13} \\
 e^{\tau  \text{ad}_{Y_{9}}} Y_{14}=e^{-\tau } Y_{14} \\
 e^{\tau  \text{ad}_{Y_{9}}} Y_{15}=e^{-\tau } Y_{15} \\
\end{array} &
\end{flalign*}

\begin{flalign*}& \begin{array}{l}
 e^{\tau  \text{ad}_{Y_{10}}} Y_1=Y_1+\tau  Y_2 \\
 e^{\tau  \text{ad}_{Y_{10}}} Y_4=Y_4+\tau  Y_5 \\
 e^{\tau  \text{ad}_{Y_{10}}} Y_6=Y_6-\tau  Y_7-\tau ^2 Y_{10} \\
 e^{\tau  \text{ad}_{Y_{10}}} Y_7=Y_7+2 \tau  Y_{10} \\
 e^{\tau  \text{ad}_{Y_{10}}} Y_8=Y_8-\tau  Y_{10} \\
 e^{\tau  \text{ad}_{Y_{10}}} Y_{11}=Y_{11}-\tau  Y_{12} \\
 e^{\tau  \text{ad}_{Y_{10}}} Y_{14}=Y_{14}-\tau  Y_{15} \\
\end{array}&
\end{flalign*}

\begin{flalign*}
&\begin{array}{l}
 e^{\tau  \text{ad}_{Y_{11}}} Y_2=Y_2+\tau  Y_3 \\
 e^{\tau  \text{ad}_{Y_{11}}} Y_4=Y_4-\tau  Y_6 \\
 e^{\tau  \text{ad}_{Y_{11}}} Y_5=Y_5-\tau  Y_8-\tau ^2 Y_{11} \\
 e^{\tau  \text{ad}_{Y_{11}}} Y_7=Y_7-\tau  Y_{11} \\
 e^{\tau  \text{ad}_{Y_{11}}} Y_8=Y_8+2 \tau  Y_{11} \\
 e^{\tau  \text{ad}_{Y_{11}}} Y_9=Y_9-\tau  Y_{11} \\
 e^{\tau  \text{ad}_{Y_{11}}} Y_{10}=Y_{10}+\tau  Y_{12} \\
 e^{\tau  \text{ad}_{Y_{11}}} Y_{13}=Y_{13}-\tau  Y_{14} \\
\end{array} &
\end{flalign*}

\begin{flalign*} &\begin{array}{l}
 e^{\tau  \text{ad}_{Y_{12}}} Y_1=Y_1+\tau  Y_3 \\
 e^{\tau  \text{ad}_{Y_{12}}} Y_4=Y_4 - \tau  \left(Y_7 + Y_8\right)-\tau ^2 Y_{12} \\
 e^{\tau  \text{ad}_{Y_{12}}} Y_5=Y_5-\tau  Y_{10} \\
 e^{\tau  \text{ad}_{Y_{12}}} Y_6=Y_6+\tau  Y_{11} \\
 e^{\tau  \text{ad}_{Y_{12}}} Y_7=Y_7+\tau  Y_{12} \\
 e^{\tau  \text{ad}_{Y_{12}}} Y_8=Y_8+\tau  Y_{12} \\
 e^{\tau  \text{ad}_{Y_{12}}} Y_9=Y_9-\tau  Y_{12} \\
 e^{\tau  \text{ad}_{Y_{12}}} Y_{13}=Y_{13}-\tau  Y_{15} \\
\end{array}&
\end{flalign*}

\begin{flalign*}
&\begin{array}{l}
 e^{\tau  \text{ad}_{Y_{13}}} Y_1=Y_1-\tau  Y_4 \\
 e^{\tau  \text{ad}_{Y_{13}}} Y_2=Y_2-\tau  Y_5 \\
 e^{\tau  \text{ad}_{Y_{13}}} Y_3=Y_3-\tau  Y_9-\tau ^2 Y_{13} \\
 e^{\tau  \text{ad}_{Y_{13}}} Y_8=Y_8-\tau  Y_{13} \\
 e^{\tau  \text{ad}_{Y_{13}}} Y_9=Y_9+2 \tau  Y_{13} \\
 e^{\tau  \text{ad}_{Y_{13}}} Y_{11}=Y_{11}+\tau  Y_{14} \\
 e^{\tau  \text{ad}_{Y_{13}}} Y_{12}=Y_{12}+\tau  Y_{15} \\
\end{array} &
\end{flalign*}

\begin{flalign*} &\begin{array}{l}
 e^{\tau  \text{ad}_{Y_{14}}} Y_1=Y_1-\tau  Y_6 \\
 e^{\tau  \text{ad}_{Y_{14}}} Y_2=Y_2 - \tau  \left(Y_8 + Y_9\right)-\tau ^2 Y_{14} \\
 e^{\tau  \text{ad}_{Y_{14}}} Y_3=Y_3-\tau  Y_{11} \\
 e^{\tau  \text{ad}_{Y_{14}}} Y_5=Y_5+\tau  Y_{13} \\
 e^{\tau  \text{ad}_{Y_{14}}} Y_7=Y_7-\tau  Y_{14} \\
 e^{\tau  \text{ad}_{Y_{14}}} Y_8=Y_8+\tau  Y_{14} \\
 e^{\tau  \text{ad}_{Y_{14}}} Y_9=Y_9+\tau  Y_{14} \\
 e^{\tau  \text{ad}_{Y_{14}}} Y_{10}=Y_{10}+\tau  Y_{15} \\
\end{array}&
\end{flalign*}

\begin{flalign*}
&\begin{array}{l}
 e^{\tau  \text{ad}_{Y_{15}}} Y_1=Y_1 - \tau  \left(Y_7 + Y_8 + Y_9\right)-\tau ^2 Y_{15} \\
 e^{\tau  \text{ad}_{Y_{15}}} Y_2=Y_2-\tau  Y_{10} \\
 e^{\tau  \text{ad}_{Y_{15}}} Y_3=Y_3-\tau  Y_{12} \\
 e^{\tau  \text{ad}_{Y_{15}}} Y_4=Y_4+\tau  Y_{13} \\
 e^{\tau  \text{ad}_{Y_{15}}} Y_6=Y_6+\tau  Y_{14} \\
 e^{\tau  \text{ad}_{Y_{15}}} Y_7=Y_7+\tau  Y_{15} \\
 e^{\tau  \text{ad}_{Y_{15}}} Y_9=Y_9+\tau  Y_{15} \\
\end{array}&
\end{flalign*}
\begin{widetext}

\section{Disentanglement equations for $\mathfrak{sl}(4,\mathbb{C})$} \label{A:disentanglement_equations}

In this Appendix we present the explicit form of the equations for $u_1, u_2, \ldots, u_{15}$ from Eq. (\ref{disentanglement_equations}).
The functions $u_1$, $u_2$, and $u_3$ satisfy the following set of three coupled Riccati equations:
\be \label{3_Riccati}
	  2  \dot{\mathbf{u}}_{(1)} = \mathbf{a}_{(1)} + C_{(1)} \mathbf{u}_{(1)} + \mathbf{u}_{(1)} \mathbf{u}_{(1)}^T \cdot \mathbf{b}_{(1)},
\ee
with
\begin{equation}
\begin{aligned}
	&\mathbf{u}_{(1)} = \left(\begin{array}{c} u_1\\u_2\\u_3
	\end{array}\right),\qquad \mathbf{a_{(1)}} = 
	\left(\begin{array}{c}
		a_1\\a_2\\a_3
	\end{array}\right),\qquad \mathbf{b}_{(1)} = 
	\left(\begin{array}{c}
	-a_{15}\\-a_{14}\\-a_{13}
	\end{array}\right),
\end{aligned}
\end{equation}
and
\be
	C_{(1)} = \left(\begin{array}{ccc} a_7 + a_9 & a_6 & a_4\\ a_{10} & -a_7 +a_8 + a_9 &a_5\\ a_{12} & a_{11} & -a_8 + 2a_9\end{array}\right).
\ee
 The functions $u_4$ and $u_5$ satisfy a set of two coupled Riccati equations:
\be \label{2_Riccati}
2\dot{\mathbf{u}}_{(2)} = \mathbf{a}_{(2)} + C_{(2)} \mathbf{u}_{(2)}  + \mathbf{u}_{(2)}\mathbf{u}_{(2)}^T \cdot \mathbf{b}_{(2)},
\ee
with
\be
\mathbf{u}_{(2)} = \left(\begin{array}{c}u_{4}\\u_{5}\end{array}\right),\qquad \mathbf{a}_{(2)} = \left(\begin{array}{c} a_4 - a_{13}u_1\\ a_5 - a_{13} u_2 \end{array}\right),\qquad \mathbf{b}_{(2)} = \left(\begin{array}{c} -a_{12} + a_{15}u_3  \\ -a_{11} + a_{14}u_3  \end{array}\right) ,
\ee
and
\be
C_{(2)} = \left(\begin{array}{cc} a_7 + a_8 - a_9 + a_{13} u_3 - a_{15} u_1 & a_6 -a_{14} u_1 \\ a_{10} - a_{15}u_2  & -a_7 + 2a_8 - a_9 - a_{14}u_2 + a_{13} u_3\end{array}\right).
\ee

The function $u_6$ satisfies a scalar Riccati equation
\be \label{1_Riccati}
\begin{aligned}
 2\dot{u}_6=&~u_6^2 \Bigl[a_{15} u_2+u_5 \left(a_{12}-a_{15} u_3\right)-a_{10}\Bigr] - a_{14}
   u_1+u_4 \left(a_{14} u_3-a_{11}\right)+a_6\\
   &+u_6 \Bigl[-a_{15} u_1+a_{14} u_2+u_4 \left(a_{15} u_3-a_{12}\right)+u_5 \left(a_{11}-a_{14} u_3\right)+2 a_7-a_8\Bigr]
\end{aligned}
\ee

Once the solutions to Eqs. (\ref{3_Riccati}), (\ref{2_Riccati}), and (\ref{1_Riccati}) are known, the remaining functions $u_7,\ldots, u_{15}$ can be found by direct integration.
Specifically, we have a set of three differential equations, completely determined by solutions of the previous six equations:
\be
\begin{aligned}
	2\dot{u}_7 &=-a_{15} u_1-a_{12} u_4+u_6\left( a_{15} u_2 +a_{12} u_5 -a_{10}\right)+u_3 \left(a_{15} u_4-a_{15} u_5 u_6\right)+a_7 ,\\
	2\dot{u}_8&=-a_{15} u_1-a_{14} u_2-a_{12} u_4-a_{11} u_5+u_3 \left(a_{15} u_4+a_{14} u_5\right)+a_8 ,\\
	2\dot{u}_9&=-a_{15} u_1-a_{14} u_2-a_{13} u_3+a_9 ;
\end{aligned}
\ee

A set of two differential equations determined by the previous nine solutions:
\be
\begin{aligned}
	2\dot{u}_{10}&=e^{2 u_7-u_8}\left(a_{10} -a_{15} u_2-u_5\left(a_{12}-a_{15} u_3 \right)\right) ,\\
	2\dot{u}_{11}&= e^{-u_7+2 u_8-u_9}\left(a_{11}+a_{12}  u_6-u_3 \left(a_{14} +a_{15} u_6\right)\right);
\end{aligned}
\ee

And finally a set of four differential equations, completely determined by the solutions to the previous eleven equations:
\be
\begin{aligned}
	2\dot{u}_{12}&=e^{-u_7+u_8-u_9} \Bigl[ e^{u_8} u_{10} \left(a_{11}+a_{12}u_6-u_3\left(a_{14} +a_{15} u_6\right) \right)+e^{2 u_7} \left(a_{12}-a_{15} u_3\right) \Bigr] ,\\
	2\dot{u}_{13}&=e^{2 u_9-u_8} \left(a_{15} u_4+a_{14} u_5+a_{13}\right) ,\\
	2\dot{u}_{14}&=e^{-u_7-u_8+u_9} \Bigl[e^{u_7+u_9}u_{11}\left(a_{13}  +a_{15}u_4+a_{14} u_5\right)+e^{2 u_8} \left(a_{15} u_6+a_{14} \right)\Bigr] ,\\
	2\dot{u}_{15}&=e^{-u_7-u_8+u_9} \Bigl[ e^{u_7+u_9}u_{12}\left(a_{13}+a_{15}u_4 +a_{14}u_5\right)+e^{2 u_8}u_{10} \left(a_{15} u_6 +a_{14}\right) +a_{15}e^{2 u_7+u_8}\Bigr].
\end{aligned}
\ee

\end{widetext}

\pagebreak


\clearpage

\begin{thebibliography}{99}

\bibitem{Lukin2017} 
H. Bernien, S. Schwartz, A. Keesling, H. Levine, A. Omran, H. Pichler, S. Choi, A.S. Zibrov, M. Endres, M. Greiner, V. Vuleti{\'c}, and M.D. Lukin,
Probing many-body dynamics on a 51-atom quantum simulator,
{\href{http://dx.doi.org/10.1038/nature24622}{Nature (London) {\bf 551}, 579 (2017)}}.

\bibitem{Monroe2017} 
J. Zhang, G. Pagano, P. W. Hess, A. Kyprianidis, P. Becker, H. Kaplan, A.V. Gorshkov, Z.-X. Gong, and C. Monroe,
Observation of a many-body dynamical phase transition with a 53-qubit quantum simulator,
{\href{http://dx.doi.org/10.1038/nature24654}{Nature (London) {\bf 551}, 601 (2017)}}.

\bibitem{Martinis2018}
C. Neill, P. Roushan, K. Kechedzhi, S. Boixo, S.V. Isakov, V. Smelyanskiy, R. Barends, B. Burkett, Y. Chen, Z. Chen, 
B. Chiaro, A. Dunsworth, A. Fowler, B. Foxen, R. Graff, E. Jeffrey, J. Kelly, E. Lucero, A. Megrant, J. Mutus, M. Neeley, 
C. Quintana, D. Sank, A. Vainsencher, J. Wenner, T.C. White, H. Neven, and J.M. Martinis,
A blueprint for demonstrating quantum supremacy with superconducting qubits,
{\href{http://dx.doi.org/10.1126/science.aao4309}{Science {\bf 360}, 195 (2018)}}.

\bibitem{Blatt2018}
N. Friis, O. Marty, C. Maier, C. Hempel, M. Holz{\"a}pfel, P. Jurcevic, M.B. Plenio, M. Huber, C. Roos, R. Blatt, and B. Lanyon,
Observation of entangled states of a fully controlled 20-qubit system,
{\href{http://dx.doi.org/10.1103/PhysRevX.8.021012}{Phys. Rev. X {\bf 8}, 021012 (2018)}}.

\bibitem{Trotzky2012}
S. Trotzky, Y.-A. Chen, A. Flesch, I.P. McCulloch, I. Schollw\"{o}ck, J. Eisert, and I. Bloch,
Probing the relaxation towards equilibrium in an isolated strongly correlated one-dimensional Bose gas,
{\href{http://dx.doi.org/10.1038/nphys2232}{Nat. Phys. {\bf 8}, 325 (2012)}}.

\bibitem{Mazurenko2017}
A. Mazurenko, C.S. Chiu, G. Ji, M.F. Parsons, M. Kan{\'a}sz-Nagy, R. Schmidt, F. Grusdt, E. Demler, D. Greif, and M. Greiner,
A cold-atom Fermi-Hubbard antiferromagnet,
{\href{http://dx.doi.org/10.1038/nature22362}{Nature (London) {\bf 545}, 462 (2017)}}.

\bibitem{Lukin2019}
A. Keesling, A. Omran, H. Levine, H. Bernien, H. Pichler, S. Choi, R. Samajdar, S. Schwartz, P. Silvi, S. Sachdev, P. Zoller, M. Endres, M. Greiner, V. Vuleti{\'c}, and M.D. Lukin,
Quantum Kibble--Zurek mechanism and critical dynamics on a programmable Rydberg simulator,
\href{https://doi.org/10.1038/s41586-019-1070-1}{Nature (London) {\bf 568}, 207 (2019)}.
	
\bibitem{Heyl2018} 
For a review, see M. Heyl, Dynamical quantum phase transitions: a review,
\href{https://doi.org/10.1088/1361-6633/aaaf9a}{Rep. Prog. Phys. {\bf 81}, 054001 (2018)}.

\bibitem{KrausZoller2008}
B. Kraus, H. P. Buchler, S. Diehl, A. Kantian, A. Micheli, and P. Zoller, 
Preparation of entangled states by quantum Markov processes,
\href{https://doi.org/10.1103/PhysRevA.78.042307}{Phys. Rev. A {\bf 78}, 042307 (2008)}.

\bibitem{Diehl2010}
S. Diehl, A. Tomadin, A.Micheli, R. Fazio, and P. Zoller, 
Quantum states and phases in driven open quantum systems with cold atoms
\href{https://doi.org/10.1103/PhysRevLett.105.015702}{Phys. Rev. Lett. {\bf 105}, 015702 (2010)}.

\bibitem{Cirac2009}
F. Verstraete, M. M. Wolf, and J. I. Cirac, 
Quantum computation and quantum-state engineering driven by dissipation,
\href{https://doi.org/10.1038/nphys1342}{Nature Phys. 5, 633 (2009)}.

\bibitem{Drummond1980} 
P. D. Drummond and D. F. Walls, 
Quantum theory of optical bistability, 
\href{https://doi.org/10.1088/0305-4470/13/2/034}{J. Phys. A: Math. Gen. {\bf13}, 725 (1980)}.
		
\bibitem{Prosen2008} 
T. Prosen and I. Pizorn, 
Quantum phase transition in a far-from-equilibrium steady state of an $XY$ spin chain.
\href{https://doi.org/10.1103/PhysRevLett.101.105701}{Phys. Rev. Lett. {\bf 101}, 105701 (2008)}.
		
\bibitem{Diehl2008} 
S. Diehl, A. Micheli, A. Kantian, B. Kraus, H. B\"uchler, and P. Zoller, 
Dynamical phase transitions and instabilities in open atomic many-body systems,
\href{https://doi.org/10.1038/nphys1073}{Nature Phys. {\bf 4}, 878 (2008)}.
	
\bibitem{DallaTorre2010}
E. G. Dalla Torre, E. Demler, T. Giamarchi, and E. Altman, 
Quantum critical states and phase transitions in the presence of non-equilibrium noise,
\href{https://doi.org/10.1038/nphys1754}{Nature Phys. {\bf 6}, 806 (2010)}.
	
\bibitem{Lindblad1976}
G. Lindblad, 
On the generators of quantum dynamical semigroups, 
\href{https://doi.org/10.1007/BF01608499}{Commun. Math. Phys. {\bf 48}, 119 (1976)}. 
		
\bibitem{Gorini1976} 
V. Gorini, A. Kossakowski and E.C.G. Sudarshan, 
Completely positive dynamical semigroups of $N$-level systems, \href{https://doi.org/10.1063/1.522979}{J. Math. Phys. {\bf 17}, 821 (1976)}.
	
\bibitem{Ringel2012}
M. Ringel and V. Gritsev, 
Liouville coherent states, 
\href{https://doi.org/10.1209/0295-5075/99/20012}{Europhys. Lett. {\bf 99}, 20012 (2012)}.

\bibitem{Ringel_2013}
M.~Ringel and V.~Gritsev, Dynamical symmetry approach to path integrals of quantum spin systems,
\href{http://dx.doi.org/10.1103/PhysRevA.88.062105}{Phys. Rev. A {\bf 88}, 062105 (2013)}.

\bibitem{Wei1963}
J.~Wei and E.~Norman, Lie algebraic solution of linear differential equations,
\href{https://doi.org/10.1063/1.1703993}{J. Math. Phys. \textbf{4}, 575 (1963)}.

\bibitem{Wei1964}
J.~Wei and E.~Norman, On global representations of the solutions of linear differential
  equations as a product of exponentials, \href{https://doi.org/10.1090/s0002-9939-1964-0160009-0}{Proc. Amer. Math. Soc. \textbf{15}, 327 (1964)}.
  
\bibitem{Charzyski2013}
S. Charzy\'nski and M. Ku\'s, 
Wei-Norman equations for a unitary evolution,
\href{https://doi.org/10.1088/1751-8113/46/26/265208}{J. Phys. A: Math. Theor. {\bf 46}, 265208 (2013)}.

%

\bibitem{Galitski_2011}
V.~Galitski, Quantum-to-classical correspondence and Hubbard-Stratonovich
  dynamical systems: A Lie-algebraic approach, \href{http://dx.doi.org/10.1103/PhysRevA.84.012118}{Phys. Rev. A {\bf 84}, 012118 (2011)}.

\bibitem{Bola_os_2015}
M.~Bola\~nos and P.~Barberis-Blostein, Algebraic solution of the Lindblad equation for a collection of
  multilevel systems coupled to independent environments. \href{http://dx.doi.org/10.1088/1751-8113/48/44/445301}{J. Phys. A: Math. Theor. {\bf 48}, 445301 (2015)}.
  
  
  \bibitem{Markovich_2017}
L.~Markovich, R.~Grimaudo, A.~Messina, and H.~Nakazato, An example of interplay between physics and mathematics: Exact resolution of a new class of Riccati equations, \href{http://dx.doi.org/10.1016/j.aop.2017.07.008}{Ann. Phys. (N.Y.), {\bf 385}, 522 (2017)}.


\bibitem{Scopa_2018} S. Scopa, G.T. Landi, and D. Karevski.
\newblock Lindblad-Floquet description of finite-time quantum heat engines. \href{http://dx.doi.org/10.1103/PhysRevA.97.062121}{Phys. Rev. A {\bf 97} (2018)}.

\bibitem{Scopa_2019} S. Scopa, G.T. Landi, A. Hammoumi, and D. Karevski.
\newblock Exact solution of time-dependent Lindblad equations with closed algebras.
\href{http://dx.doi.org/10.1103/PhysRevA.99.022105}{Phys. Rev. A, \textbf{14} (2019).}



\bibitem{De_Nicola_2019}
S.~De~Nicola, B.~Doyon, and M.~J. Bhaseen, Stochastic approach to non-equilibrium quantum spin systems.
\href{http://dx.doi.org/10.1088/1751-8121/aaf9be}{J. Phys. A: Math. Theor. {\bf 52}, 05LT02 (2019)}.

\bibitem{De_Nicola_2020}
S.~De~Nicola, B.~Doyon, and M.~J. Bhaseen, Non-equilibrium quantum spin dynamics from classical stochastic processes, \href{http://dx.doi.org/10.1088/1742-5468/ab6093}{J. Stat. Mech. {\bf 2020}, 013106 (2020)}.

\bibitem{Vernier2020}
E. Vernier, Mixing times and cutoffs in open quadratic fermionic systems, \href{http://dx.doi.org/10.21468/SciPostPhys.9.4.049}{SciPost Phys. {\bf 9}, 049 (2020)}.
%
\bibitem{Zoller2011}
S. Diehl, E. Rico, M.A. Baranov, and P. Zoller, 
Topology by dissipation in atomic quantum wires, 
\href{https://doi.org/10.1038/nphys2106}{Nature Phys. {\bf 7}, 971 (2011)}.

\bibitem{Baranov2012}
C.-E. Bardyn, M. A. Baranov, E. Rico, A. Imamoglu, P. Zoller, and S. Diehl,
Majorana modes in driven-dissipative atomic superfluids with a zero Chern number, 
\href{https://doi.org/10.1103/PhysRevLett.109.130402}{Phys. Rev. Lett. {\bf 109}, 130402 (2012)}.

\bibitem{Zoller2010}
S. Diehl, W. Yi, A. Daley, and P. Zoller, 
Dissipation-induced $d$-wave pairing of fermionic atoms in an optical lattice, 
\href{https://doi.org/10.1103/PhysRevLett.105.227001}{Phys. Rev. Lett. {\bf 105}, 227001 (2010)}.

\bibitem{Zoller2012}
W. Yi, S. Diehl, A. Daley, and P. Zoller, 
Driven-dissipative many-body pairing states for cold fermionic atoms in an optical lattice, 
\href{https://doi.org/10.1088/1367-2630/14/5/055002}{New J. Phys. {\bf 14}, 055002 (2012)}.

\bibitem{Keeling2012}
F. Nissen, S. Schmidt, M. Biondi, G. Blatter, H. E. T\"ureci, and J. Keeling, 
Nonequilibrium dynamics of coupled qubit-cavity arrays, 
\href{https://doi.org/10.1103/PhysRevLett.108.233603}{Phys. Rev. Lett. {\bf 108}, 233603 (2012)}.

\bibitem{Kitaev2001}
A.Y. Kitaev, 
Unpaired Majorana fermions in quantum wires, 
\href{https://doi.org/10.1070/1063-7869/44/10S/S29}{Phys. Usp. {\bf 44}, 131 (2001)}.

\bibitem{proof}
Indeed, according to Eq. (\ref{Lambda_k}), the only way one can have $\Lambda_k = 0$ is when $u_k = v_k = 0$. The latter means that on this momentum mode $k$ there is no dissipation at all [see Eq. (\ref{Jump_op_k_space})]. Hence, the evolution is unitary and $\Re{\lambda_j} = 0$ for all $j$. This also follows directly from Eq. (\ref{spectrum}) since for $u_k = v_k = 0$ we have $\Theta_k = \Phi_k = 0$, which leads to $U_k<0$ and $V_k = 0$.

\end{thebibliography}
\end{document}